\documentclass[useAMS,usenatbib]{mn2e}

\usepackage{graphicx,amssymb,color}
\usepackage[normalem]{ulem}
\usepackage{longtable}

\title[Asteroid disruption sims for WD 1145+017]
{Explaining the variability of WD 1145+017 with simulations of asteroid tidal disruption}

\author[Veras, Carter, Leinhardt \& G\"{a}nsicke]{
Dimitri Veras$^{1}$\thanks{E-mail: d.veras@warwick.ac.uk},
Philip J. Carter$^{2}$,
Zo\"{e} M. Leinhardt$^{2}$,
Boris T. G\"{a}nsicke$^{1}$
\\
$^{1}$Department of Physics, University of Warwick, Coventry CV4 7AL, UK
\\
$^{2}$School of Physics, University of Bristol, Bristol BS8 1TL, UK
}

\date{Accepted 2016 October 21. Received 2016 October 21; in original form 2016 July 20}

\begin{document}
\label{firstpage}
\pagerange{\pageref{firstpage}--\pageref{lastpage}} \pubyear{2016}
\maketitle

\begin{abstract}
Post-main-sequence planetary science has been galvanised
by the striking variability, depth and shape of the photometric transit curves
due to objects orbiting white dwarf WD 1145+017, a star which
also hosts a dusty debris disc and circumstellar gas, and displays strong metal atmospheric
pollution. However, the physical properties of the likely asteroid which is 
discharging disintegrating fragments remain largely unconstrained
from the observations. This process has not yet been modelled
numerically. Here, we use the $N$-body code {\tt PKDGRAV} to compute dissipation
properties for asteroids of different spins, densities, masses, and eccentricities.
We simulate both
homogeneous and differentiated asteroids, for up to two years, and find that the disruption timescale
is strongly dependent on density and eccentricity, but weakly dependent on mass and spin.
We find that primarily rocky differentiated bodies with moderate 
($\sim 3-4$ g/cm$^3$)
bulk densities on near-circular ($e \lesssim 0.1$) orbits can remain intact while 
occasionally shedding mass from their mantles.
These results suggest that the asteroid orbiting WD 1145+017 is differentiated,
resides just outside
of the Roche radius for bulk density but just inside the Roche radius for mantle density, 
and is more akin physically to an asteroid like Vesta instead of one like Itokawa.
\end{abstract}


\section{Introduction}

Observations of the fates of planetary systems help constrain their formation and subsequent
evolution, and provide unique insights into their bulk composition.  Planets, moons and
asteroids which survive engulfment from their parent star's giant branch evolution
\citep{villiv2009,kunetal2011,musvil2012,adablo2013,norspi2013,viletal2014,payetal2016a,
payetal2016b,staetal2016}
represent a sufficient reservoir of material to eventually ``pollute'' between
one-quarter and one-half of all Milky Way white dwarfs with metals
\citep{zucetal2003,zucetal2010,koeetal2014}. This fraction is roughly commensurate
with that of planet-hosting main sequence stars \citep{casetal2012}.

The high mass density of white dwarfs ($\sim$~$10^5$~$-$~$10^6$~g~cm$^{-3}$) ensures 
that their atmospheres stratify chemical elements \citep{schatzman1958}, allowing for the relatively
easy detection of metals \citep{zucetal2007,kleetal2010,kleetal2011}, particularly with 
high-resolution ultraviolet spectroscopy \citep{xuetal2013,xuetal2014,wiletal2015,wiletal2016}.
Consequently, trends amongst the chemical diversity and bulk composition of exoasteroids,
which are the building blocks of planets, may be inferred and linked to specific
families in the Solar system \citep{ganetal2012,juryou2014} or to the
compositional evolution during accretion of exoplanets themselves \citep{caretal2015}.  

The pollutants are accreted from either or both surrounding debris discs
and direct impacts. About 40 white dwarf debris discs have now been identified
\citep{zucbec1987,becetal2005,gaeetal2006,gaeetal2008,faretal2009,
baretal2012,wiletal2014,farihi2016,manetal2016}, exclusively around
white dwarfs which are polluted, strengthening the link between pollution and debris
discs. Bodies may frequently impact the white dwarf directly \citep{wyaetal2014,broetal2016},
including comets \citep{alcetal1986,veretal2014a,stoetal2015}, moons 
\citep{payetal2016a,payetal2016b}, asteroids \citep{bonetal2011,debetal2012,frehan2014,antver2016},
or small planets \citep{hampor2016}.
Alternatively, upon entering the Roche (or disruption) radius, one of these bodies
may break up, forming a disc 
\citep{graetal1990,jura2003,debetal2012,beasok2013,veretal2014b,veretal2015a}
which eventually, and in a nontrivial manner, accretes onto the white dwarf
\citep{bocraf2011,rafikov2011a,rafikov2011b,metetal2012,rafgar2012}.

How planets might perturb these smaller bodies into the Roche radius is a growing
field of study \citep{veras2016a} that is now buttressed with self-consistent simulations
merging stellar evolution and multi-planet dynamics over many
Gyr \citep{veretal2013a,musetal2014,veras2016b} 
or even over one Hubble time \citep{vergae2015,veretal2016a,veretal2016c}.
Planets are generally required as perturbing agents because
self-perturbation into the Roche 
radius due to radiative effects alone is unlikely \citep{veretal2015b,veretal2015c}.
Amongst the presence of external stellar perturbers, planets provide
a key pathway for smaller bodies to collide with the white dwarf
\citep{bonwya2012,bonver2015,petmun2016}.

Before the year 2015, what was missing from the framework detailed 
above were detections of asteroids breaking up within the Roche radius of a white 
dwarf. That situation changed with the discovery of photometric transits from
\textit{K2} light curves of WD 1145+017. These strongly suggest that at least one body 
around this white dwarf is disintegrating \citep{vanetal2015}. The transit signatures change shape
and depth on a nightly basis \citep{gaeetal2016,rapetal2016} in a manner that is
unique amongst exoplanetary systems, prompting intense follow up studies 
\citep{croetal2015,aloetal2016,garetal2016,redetal2016,xuetal2016,zhoetal2016}.
A plausible interpretation of the observations, exemplified by fig. 7 of \cite{rapetal2016},
is that a single asteroid is disintegrating and producing multiple nearly
co-orbital fragments. However, the actual tidal disruption has not yet been modelled numerically, 
and, with the exception of \cite{guretal2016}, all previous studies on this system have been 
observationally-focused.

In this paper, we perform this task, and utilize both homogeneous and differentiated
rubble piles to model the evolution of an object which could create the observational 
transit signatures. We first, in Section 2, describe the known parameters
of the objects orbiting WD 1145+017. Then, in order to begin quantifying disruption, in 
Section 3 we summarize different 
simple formulations of the Roche radius that have appeared in the literature and
how they relate to our simulations. The setup for these simulations is described
in Section 4, and the results are reported in Sections 5 and 6.  We 
discuss the implications for WD 1145+017 and utility of our study to similar
systems in Section 7, and conclude in Section 8.

\section{Known parameters}

The known orbital parameters of the WD 1145+017 system are effectively limited
to the orbital periods of individual transit features.  These periods are 
obtained directly from the photometric transit curves for features
which are observed over multiple nights, and are known to exquisite
precision. As suggested by \cite{rapetal2016},
only one periodic signal, at 4.50 hours, has been consistently detected
over a timespan of about two years 
(going back to the \textit{K2} observations presented by \citealt*{vanetal2015})
and may be associated with the major disrupting parent body.
Other transit features, which could be interpreted as co-orbital disintegrating debris,
have orbital periods ranging from 4.490 
to 4.495 hours \citep{gaeetal2016,rapetal2016}.

These periods, however, do not translate into well-constrained semimajor axes 
because the stellar mass, $M_{\rm WD}$, is not well known.  However, under the reasonable 
assumption \citep[e.g.][]{treetal2016} that
$M_{\rm WD}$~$=$~$0.5M_{\odot}$~$-$~$0.7M_{\odot}$, the semimajor axis $a$ of the 
disrupting asteroid -- henceforth
denoted as the parent body -- lies in the range $0.0051-0.0057$ au 
(assuming a negligible-mass parent body). The typically-used 
fiducial white dwarf mass of $M_{\rm WD} = 0.6M_{\odot}$ gives $a = 0.0054$ au. 

\section{Roche radius}

Does this semimajor axis value lie within the white dwarf Roche radius?
The answer is not obvious because it depends on how the Roche radius is defined  
(see equation 9.1 of \citealt*{veras2016a} for a summary of the different equations in 
the post-main-sequence literature)
and how much internal strength the bodies are assumed to have \citep{corsha2008,veretal2014a,beasok2015}. 
We will ignore material strength for the remainder of this paper because for objects larger than about
10 km, the gravitational binding energy is more significant than any material strength
\citep{benasp1999,leiste2012}.

Assume that a Roche radius $r_{\rm d}$ defines a disruption sphere. Hence

\begin{equation}
r_{\rm d} \propto R_{\rm WD} \left( \frac{\rho_{\rm WD}}{\rho} \right)^{1/3} 
         \propto          \left( \frac{M_{\rm WD}}{\rho} \right)^{1/3}
\label{eqfirst}
\end{equation}

\begin{equation}
 \ \ \ \ = K R_{\rm WD} \left( \frac{\rho_{\rm WD}}{\rho} \right)^{1/3} 
         = k \left( \frac{M_{\rm WD}}{\rho} \right)^{1/3}
\label{rochekK}
\end{equation}

\noindent{}where $\rho$ is the bulk density of the orbiting body, and $\rho_{\rm WD}$ and $R_{\rm WD}$ are
the density and radius of the white dwarf. The definition ambiguity arises
with the choice of proportionality constants, $K$ or $k$, which are related through
$K \approx 1.61 k$.

The exact value of $K$ (or $k$) depends on the nature of the body; see Table \ref{RocheTable}.
A more thorough treatment in the strengthless
case \citep{davidsson1999} reveals that $r_{\rm d}$ should also be explicitly dependent on
the body's tensile strength and shear strength, parameters which determine when the body will specifically 
fracture or split.

A useful Roche radius expression that is rescaled for white dwarf systems is \citep[Eq. 1 of][]{beasok2013}

\begin{equation}
\frac{r_{\rm d}}{R_{\odot}} = 0.65 C \left( \frac{M_{\rm WD}}{0.6 M_{\odot}} \right)^{1/3}
                                   \left( \frac{\rho}{3 \ {\rm g \, cm}^{-3} } \right)^{-1/3}
\end{equation}

\noindent{}which is related to equation (\ref{rochekK}) through $C = 1.63 k$. \cite{beasok2013}
stated that $C = 1.3-2.9$, which hence corresponds to $k = 0.80-1.78$ and $K = 1.29-2.87$.
Motivated by Chandrasekhar's seminal work \citep{chandrasekhar1969}, \cite{leietal2012} instead
defined a proportionality constant $\mathcal{R}$ in their equation 5 as, 

\begin{equation}
\mathcal{R} \equiv \frac{3}{4} \frac{\rho}{\rho_{\rm WD}} \left( \frac{r_{\rm d}}{R_{\rm WD}} \right)^3
\label{zoec}
\end{equation}

\noindent{}which can be shown to be related to $K$ through
$\mathcal{R} = 0.75 K^3$ (see Table \ref{RocheTable}). Overall, $K$ varies
only within a factor of two for these different scenarios.

\begin{table}
\caption{Proportionality constants for different formulations (Equations \ref{rochekK}-\ref{zoec})
of the Roche radius and different generalized strengthless
body types. The relations between constants are $K = 1.61k$, $C = 1.63k$, and $\mathcal{R} = 0.75K^3$.
The term ``spinning'' assumes synchronous spinning from being tidally synchronized.
\newline 
Notes:
     \newline $^{\rm a}$ See equation 4.131 of Murray \& Dermott (1999).
     \newline $^{\rm b}$ Sridhar \& Tremaine (1992)
     \newline $^{\rm c}$ Roche (1847)
}
\begin{tabular}{c  c  c  c  c  c}
\hline
body type  & $K$ & $k$ & $C$ & $\mathcal{R}$ & Ref. \\
\hline
solid no spin  & 1.26 & 0.78 & 1.27 & 1.50 &   \\
solid spinning    & 1.44 & 0.89 & 1.45 & 2.24 & a \\
fluid no spin  & 1.69 & 1.05 & 1.70 & 3.62 & b \\
fluid spinning    & 2.46 & 1.53 & 2.49 & 11.2 & c \\
\hline
\end{tabular}
\label{RocheTable}
\end{table}

Constraining the numerical values of these constants enables one to estimate
$\rho$ from knowledge of the orbital period $P$ alone. If one assumes that
the body's orbit is circular
and the body resides just at the Roche radius, then comparing equation \ref{rochekK}
with Kepler's third law cancels out the (unknown) white dwarf mass and leads to the
relation

\begin{eqnarray}
\rho &=& \frac{4 \pi^2 k^3}{G P^2} 
\nonumber
\\
     &=& \left(0.72 {\rm \ g \, cm}^{-3}\right) \mathcal{R} \left( \frac{P}{4.5 \, {\rm hours}} \right)^{-2}
.
\label{denper}
\end{eqnarray}

\begin{figure}
\centerline{
\includegraphics[width=8.7cm]{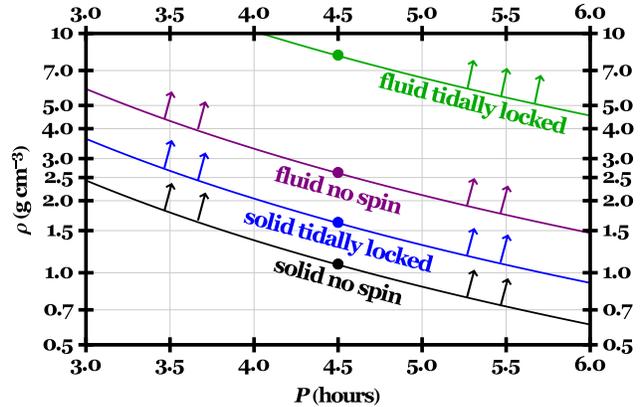}
}
\caption{
The relation between density ($\rho$) and orbital period ($P$) of a strengthless spherical
homogeneous rubble pile which resides just at the Roche radius (equation \ref{denper}); tidally locked rubble
piles are in synchronous rotation and arrows indicate stable regions. The relation is general, 
and is independent of the properties of the central object. Dots indicate the 
orbital period of the likely disintegrating asteroid at WD 1145+017.
}
\label{dengrid}
\end{figure}

Equation (\ref{denper}) provides a useful quick way to estimate the density of
a disrupting body if only its orbital period is known. For the bodies orbiting WD 1145+017
(with periods of roughly 4.5 hours),
this formula gives $\rho = 1.08, 1.62, 2.61, 8.04$ g cm$^{-3}$ for the solid 
no spin, solid spinning, fluid no spin, and fluid spinning, cases, respectively. 
We plot equation (\ref{denper}) in Fig. \ref{dengrid} for the purpose of wider use beyond 
this individual planetary system.

All of the bodies in the figure are strengthless. Incorporating strength would complicate equations (\ref{eqfirst}-\ref{zoec}), and the resulting curves would be different.  The fluid case is not important for the numerical aspects of our study, but was included in Fig. \ref{dengrid} and Table \ref{RocheTable} for comparison with and clarification of existing literature on Roche radii formulations. In contrast to fluids, granular materials, in general, can withstand considerable shear stress when they are under pressure \citep[Sec. 4.2 of][]{manetal2009}.

%%%%%%%%%%%%%%%%% Figure Snapshot
%%%%%%%%%%%%%%%%% Figure Snapshot

\section{Numerical methods}

The analysis from the last section shows that knowledge of $P$ can provide strong
constraints on $\rho$. We explore this possibility in the case of WD 1145+017 with numerical simulations
of rubble piles, which are aggregates bound together by self-gravity.
We used the $N$-body gravity tree code \textsc{PKDGRAV} \citep{stadel2001}, which has been modified with the
ability to detect and resolve collisions amongst individual particles \citep{leietal2000,ricetal2000}.

\subsection{Common properties}

Our simulations required us to specify a central mass and semimajor axis. 
We chose values that correspond to $P\approx$~4.5\,hours ($M_{\odot}$~$=$~$0.60M_{\odot}$ 
and $a$~$=$~$0.0054$ au)
for all simulations because our focus is on modelling the WD 1145+017 system.  
We adopted a constant timestep of 50 seconds for all simulations, a choice which is sufficient to 
resolve the collisions within each rubble pile (see Section 5.3 of \citealt*{veretal2014b} for details).

\subsection{Other parameter choices}

Other parameters that we varied amongst different simulations include
rubble pile structure, number of particles, bulk density $\rho$ and mass $M$ (and hence radius $R$), 
plus eccentricity $e$ and spin. See Table \ref{simtable} for the full list of simulations; three highly-referenced
simulations are repeated here in Table \ref{ShortTab} for demonstration purposes. The table columns are

\begin{figure}
\centerline{
\includegraphics[width=8.5cm]{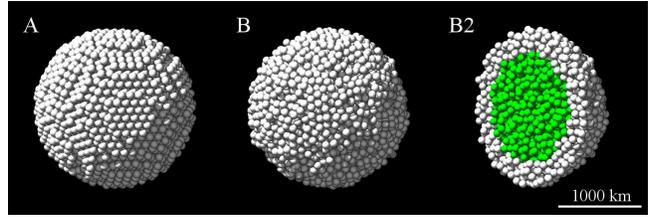}
}
\caption{
({\it A:}) A rubble pile with hexagonal closest packing (HCP) of 5003 particles.  
({\it B:}) A randomly-packed differentiated rubble pile consisting of 5000 particles. 
({\it B2:}) A hemispherical cutaway of {\it B} to show the structure (not a distinct initial condition).
}
\label{rubpile}
\end{figure}

\begin{itemize}

\item {\bf packing type}: We created our rubble piles with two 
different internal packing structures: 1) hexagonal closest packing \citep[HCP;][]{leietal2000} and 
2) random packing.  See Fig. \ref{rubpile} for a visual comparison.

\item {\bf differentiated}: This column indicates if the rubble pile was homogeneous
or differentiated. The differentiated rubble piles all contained
a ``core'' (green particles in Fig. \ref{rubpile}; image B2) and ``mantle'' 
(white particles in Fig. \ref{rubpile}; image B2). 
Each type of particle has different properties: each core particle was four times more massive than 
each mantle particle, although all particles had the same size. For these rubble piles, about 
35 per cent of particles were core particles.

\item {\bf number of particles}: The vast majority of our simulations contained about 5000 particles, which
is a well-justified choice \citep[e.g.][]{leietal2012,veretal2014b} because disruption properties have been shown to be independent of particle number until it becomes smaller than roughly 1000. Some long-duration simulations necessarily featured smaller number of particles due to computational limitations.

\item {\bf density}: Our choices for $\rho$ ($1.0-4.6$ g cm$^{-3}$) were motivated by 
Fig.~\ref{dengrid} and encompassed the Roche radius regions for solid parent bodies which may
be spinning, and a fluid parent body with no spin. This range is reasonable but not exhaustive: 
the differentiated bodies limit the density on the upper end because the density of the core would 
be too high if the bulk density was much above 4.6.

\item {\bf mass}: We have considered different parent body masses, even though theoretically tidal disruption of a strengthless body should be scale-independent for any asteroid size (and therefore be independent of mass for a set density; see \citealt*{solem1994}).  We sampled seven orders of magnitude in parent body mass ($M = 10^{16}-10^{22}$ kg), a range which is bounded from below by parent bodies whose mutual co-orbital interactions would produce period 
variations on the order of tenths of seconds \citep{guretal2016} and from above based on 
when instability might set in at a significant level \citep{veretal2016b}.  Planet-mass objects are 
not assumed to frequently enter white dwarf Roche radii \citep{veretal2013a,musetal2014,vergae2015,veretal2016a,veras2016b}
unless they are smaller than the terrestrial planets and, perhaps, are perturbed by a stellar companion 
\citep{hampor2016,petmun2016}. 

\item {\bf radius}: The radius was simply computed from our choices of $M$ and $\rho$.

\item {\bf $e$ (eccentricity)}:  
Observations so far do not constrain eccentricity, and in the absence of constraints, circular 
orbits are the simplest assumption. We sampled eccentricities up to 0.2; see Section \ref{ressec}
for more details.

\item {\bf spin}: The values of {\it 0}, {\it 1} and {\it 2} indicate no spin,
synchronous spin, and twice the synchronous spin rate.  A spin value of 0 effectively refers
to rotation once per orbit in the direction opposite to the motion in the corotating frame. 
A spin value of 2 instead refers to rotation once per orbit in the orbit direction in the 
corotating frame.

\item {\bf duration}: All simulations were run for at least three months (90 days), and
some up to two years (730 days).  The timespan of 90 days is both computationally
feasible and observationally motivated (WD 1145+017 is observed on an almost nightly basis
and hence well-sampled over the course of months).  This timespan covered nearly 481 orbits. 
Two years corresponds to about 3900 orbits, which represents the 
overall baseline of observations of the disintegrating asteroid, dating back to the 
{\it K2} observations reported by \cite{vanetal2015}.

\item {\bf outcome}:  The homogeneous cases result in either {\em full} or no disruption ({\em none}).
The differentiated primary bodies could result in one additional outcome: {\em mantle} disruption,
where some mantle is lost but the core remains intact.

\end{itemize}

Overall, computational limitations required us to judiciously choose the resolution
with which to sample each parameter range, and how to partition our choices amongst
different rubble pile constructions.

%%%%%%%%%%%%%%%%% Figure Snapshot
%%%%%%%%%%%%%%%%% Figure Snapshot

%%%%%%%%%%%%%%%%% Figure Heat
\begin{figure}
\centerline{
\includegraphics[width=8.7cm]{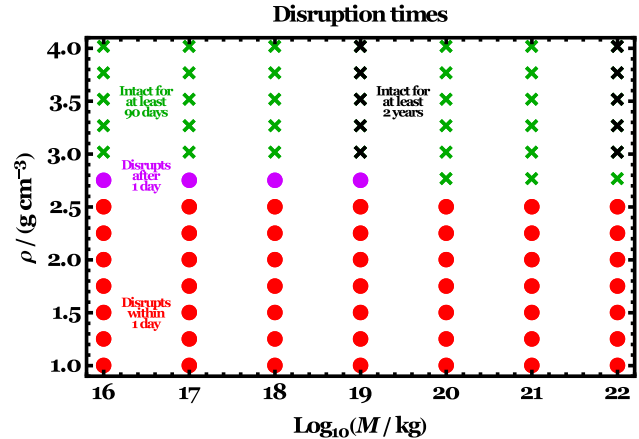}
}
\caption{Disruption characteristics for homogeneous HCP progenitors as a function of
parent body mass ($M$) and density ($\rho$) for circular orbits.  
Dots illustrate rubble piles which 
disrupted ({\tt HCP1} to {\tt HCP53}) within one
day (red dots) or between one day and one month (purple dots).
Crosses represent rubble piles that
remained stable throughout their simulations ({\tt HCP54} to {\tt HCP111}), with 
durations of 90 days (green crosses) or 2 years (black crosses). 
%All simulations listed in Tables \ref{SimSummary1}-\ref{SimSummary2} are presented
%on this plot. 
The density boundary between disruption and remaining intact is sharp,
and is between 2.5 and 3.0 g cm$^{-3}$ for all masses sampled.
}
\label{heat}
\end{figure}
%%%%%%%%%%%%%%%%% Figure Heat

\begin{table*}
\caption{Details of three simulations which appear throughout the text and in Figs. \ref{homodis}-\ref{transit}. 
All simulations performed in this work are detailed in the Appendix.}
\begin{tabular}{| c  c  c  c  c  c  c  c  c  c  c}
\hline
simulation &  packing  &   differe-    & number of  & density      & mass      & radius     & $e$ & spin & duration & outcome \\
   name    &   type    &   ntiated     &  particles &  (g cm$^{-3}$) & (kg)      &  (km)      &            &      &  (days)  & (disruption type) \\
\hline
HCP134    &  hexagonal &        no  & 5003   &  2.60     & $1.0 \times 10^{22}$  & $1000$   & 0.00         & 0 & 90  &  full  \\
RandDiff19   &  random  &        yes  & 5000   &  3.50      & $1.2 \times 10^{22}$  & $1000$   & 0.00        & 1 & 90  &   mantle  \\
RandDiff32    &  random  &        yes  & 5000   &  3.50     & $1.2 \times 10^{22}$  & $1000$   & 0.01         & 1 & 90  &  full    \\
\hline
\end{tabular}
\label{ShortTab}
\end{table*}

%%%%%%%%%%%%%%%%% Figure Ecc
\begin{figure*}
\centerline{
\includegraphics[width=8.7cm]{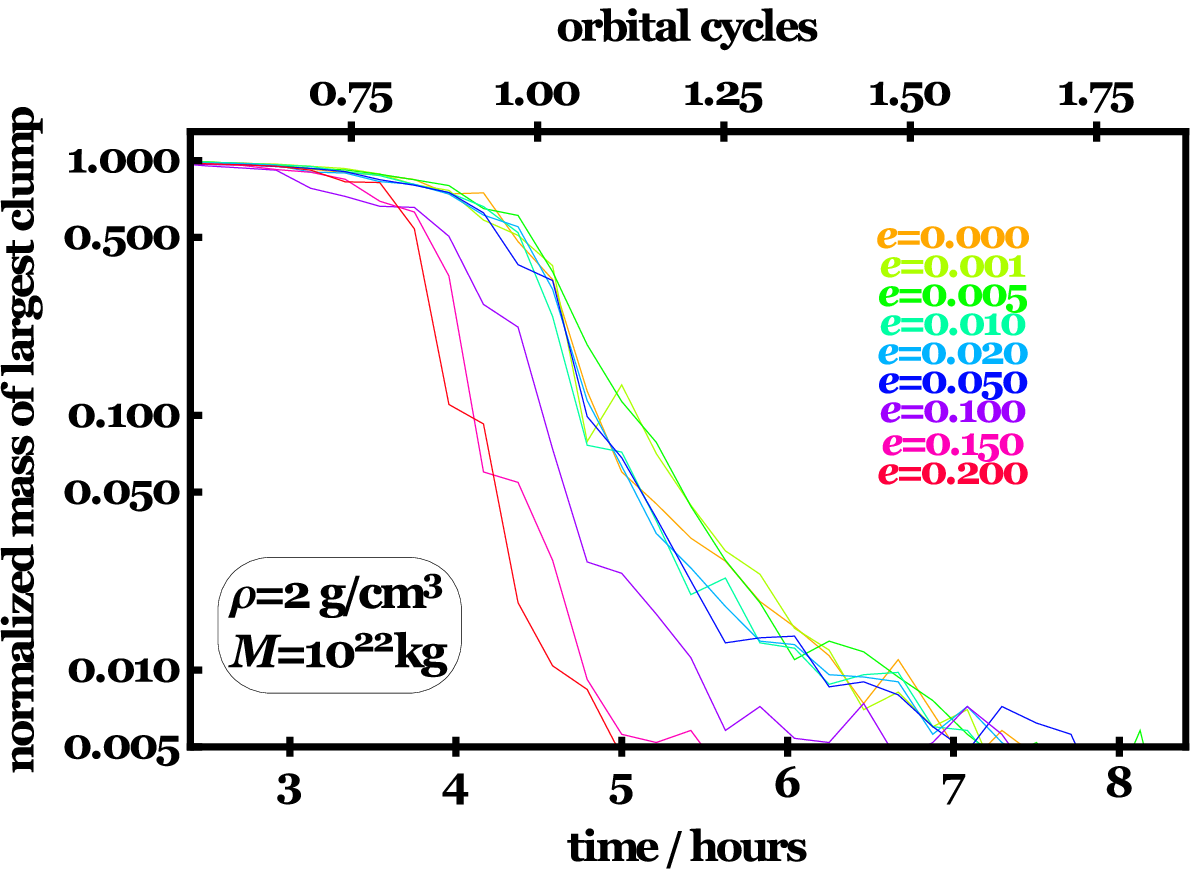}
\
\
\includegraphics[width=8.7cm]{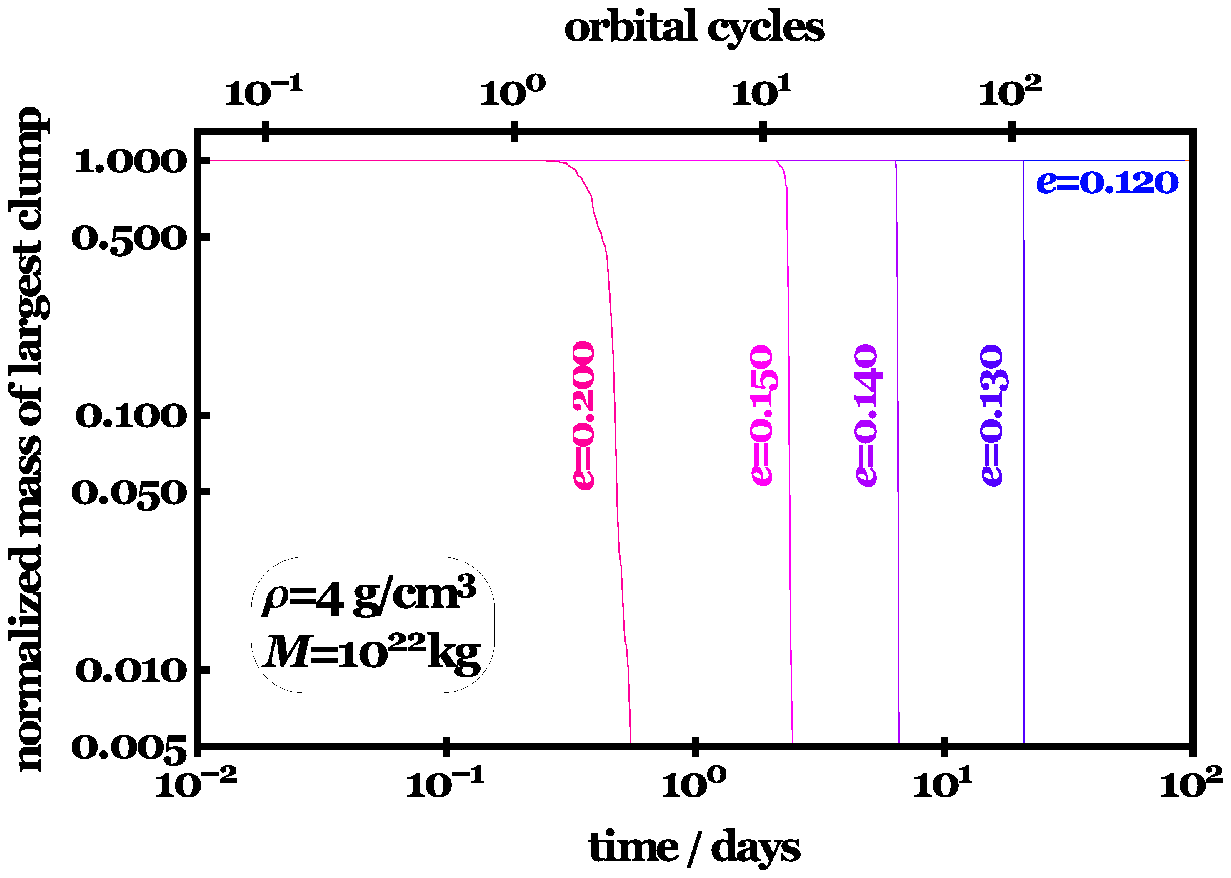}
}
\caption{
Disruption timescales for $M=10^{22}$ kg, different eccentricities (shown on plots), 
and two different parent body densities ($2$\,g\,cm$^{-3}$, left panel; $4$\,g\,cm$^{-3}$, right 
panel). Included on the figure are all simulations labeled {\tt HCP112} through {\tt HCP133} 
from Table \ref{simtable} (the horizontal curves for {\tt HCP120} to {\tt HCP129} in the right
panel all overlap). In the left panel, all
parent bodies fully disrupt within two orbits, regardless of eccentricity; generally, the 
higher the eccentricity, the quicker the dissipation. The right panel demonstrates this
correlation clearly (a disruption timescale of three months occurs for some $e$ value
between 0.12 and 0.13).
}
\label{Eccentricity}
\end{figure*}
%%%%%%%%%%%%%%%%% Figure Ecc

\section{Simulation results} \label{ressec}

An observationally relevant question is, can we match the transit observations with a disrupting rubble pile?

\subsection{Homogeneous rubble piles}

\begin{figure*}
\centerline{
\includegraphics[width=18.0cm]{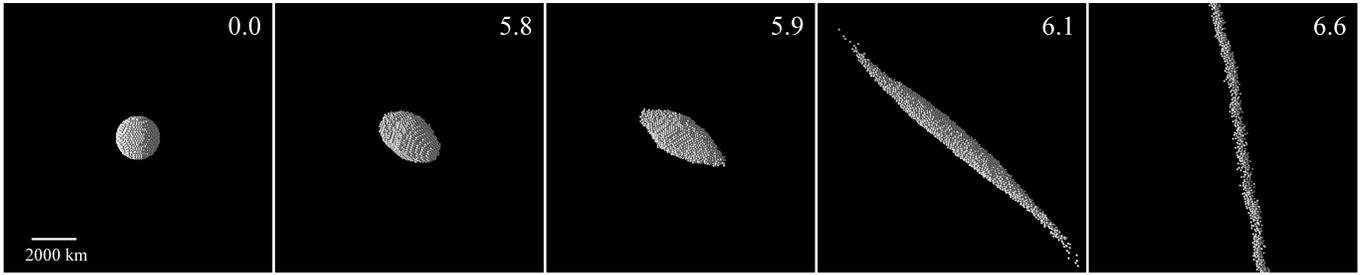}
}
\caption{
The disruption of a homogeneous hexagonal closest packing (HCP) rubble pile from 
simulation {\tt HCP134} ($\rho = 2.6$ g cm$^{-3}$, $e=0$). The images are shown 
in the rotating frame, where left is radially towards the white dwarf and the 
direction of the orbit is towards the top of the page. The white numbers
in the upper part of each panel refer to the number of orbits. An animation
accompanying this figure is available online.
}
\label{homodis}
\end{figure*}

Our first attempt to tackle the answer is to consider the simplest object: a homogeneous one.
Also, one of the strictest observational constraints is that the transits are still observed after 2 years.
Therefore, this constraint is the first which we try to replicate. We do so by presenting our results primarily in terms of
how quickly the parent body fully disintegrates. We consider a rubble pile to have disrupted after the mass of the most massive
remaining clump is less than one per cent of mass of the original rubble pile. This disruption process
for homogeneous rubble piles is illustrated in Fig. \ref{homodis}.

\subsubsection{Density constraints}

Figure \ref{heat} demonstrates disruption times as a function of
$M$ and $\rho$ (for simulation details see Table \ref{simtable})
for rubble piles on circular orbits.
The dots consist of all rubble piles which disrupted within 90 days;
nearly all of these are red dots, indicating disruption within one day.
Alternatively, the crosses signify rubble piles that remained intact
throughout the duration of the simulation; green crosses represent
our standard numerical resolution 90-day simulations and black crosses represent 
lower numerical resolution 2-year simulations. 
The boundary between the dots and crosses, at $\rho \approx 2.5-3.0$ g~cm$^{-3}$,
is sharp. The plot
illustrates the strong sensitivity of disruption to this density boundary, 
and the relative insensitivity to mass.

\subsubsection{Eccentricity constraints}

The last section demonstrates that homogeneous primary bodies on circular orbits either disrupt quickly
or not at all. These two possbilities do not aid in the interpretation of the observations.
Therefore, here we consider non-zero eccentricities. These might allow a rubble pile to dip
in-and-out of the Roche radius, shedding some mass during every pericentre passage.

Figure \ref{Eccentricity} displays results for simulations of non-circular
rubble piles.  The simulations in the figure utilise the same semimajor axis as the zero-eccentricity simulations, so that the variations seen are strongly dependent on the decreasing periapse at increasing eccentricity. The plots suggest that for a sufficiently high bulk density
($\rho \gtrsim 2.5$ g\,cm$^{-3}$ from Fig. \ref{heat}), there exists a critical 
eccentricity below which the parent body will remain intact for at least three months.  
For $\rho = 2$ g\,cm$^{-3}$ (left panel) rubble piles disrupt even on circular orbits, 
and increasing eccentricity speeds up the disruption (from about 8 hours to 5 hours).
In the
right panel, where  $\rho = 4$ g\,cm$^{-3}$, this critical eccentricity lay
in-between 0.12 and 0.13. Further, the right panel indicates a clearly monotonic
trend between increasing eccentricity and disruption time: $e=0.20$ corresponds to disruption
within a day, whereas rubble piles with $0.14$~$\le$~$e$~$\le$~$0.15$ disrupt within a week,
and those with $e = 0.13$ take about a month to break apart.

%%%%%%%%%%%%%%%%% Figure Ecc
%%%%%%%%%%%%%%%%% Figure Ecc

%%%%%%%%%%%%%%%%% Figure Ecc
\begin{figure*}
\centerline{
\includegraphics[width=18.0cm]{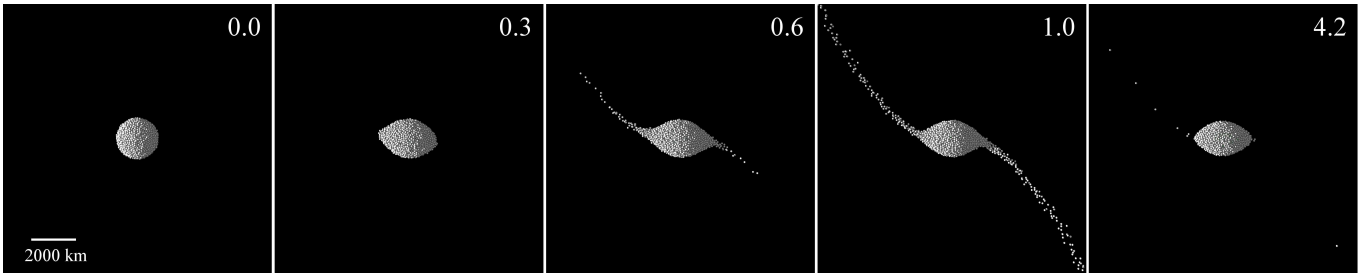}
}
\caption{
Mantle disruption of a differentiated synchronously-spinning
rubble pile on a circular orbit with $\rho = 3.5$ g \,cm$^{-3}$
(simulation {\tt RandDiff19}). 
The white particles are mantle particles, and the green core
particles underneath remain hidden. After about half of an orbit,
mantle particles start streaming from the L1 and L2 Lagrange points.
After about four orbits, the streaming became intermittent. An animation
accompanying this figure is available online.
}
\label{mandis}
\end{figure*}
%%%%%%%%%%%%%%%%% Figure Ecc

%%%%%%%%%%%%%%%%% Figure Ecc
\begin{figure*}
\centerline{
\includegraphics[width=18.0cm]{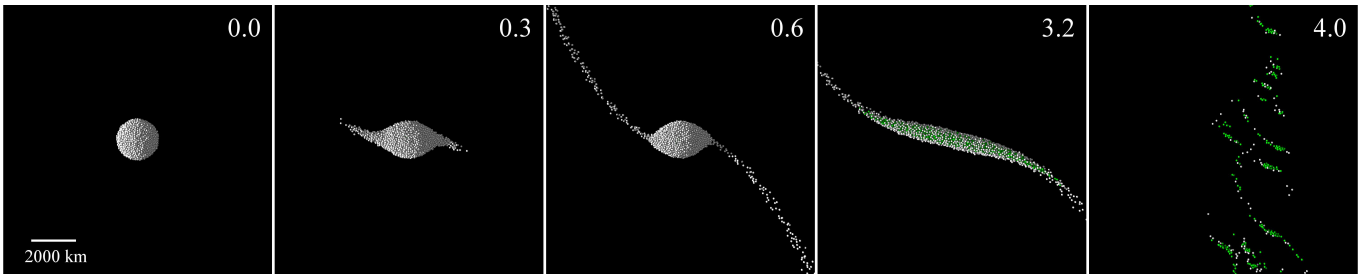}
}
\caption{
Similar to Fig. \ref{mandis}, except for the case of complete
disruption with $e=0.10$ (simulation {\tt RandDiff32}). 
Subsequent to mantle stripping, the core is not dense enough to resist disruption, and
both the white mantle particles and green core particles are visible after three orbits.
An animation
accompanying this figure is available online.
}
\label{fulldis}
\end{figure*}
%%%%%%%%%%%%%%%%% Figure Ecc

%%%%%%%%%%%%%%%%% Figure Ecc
\begin{figure*}
\centerline{
\includegraphics[width=18.0cm]{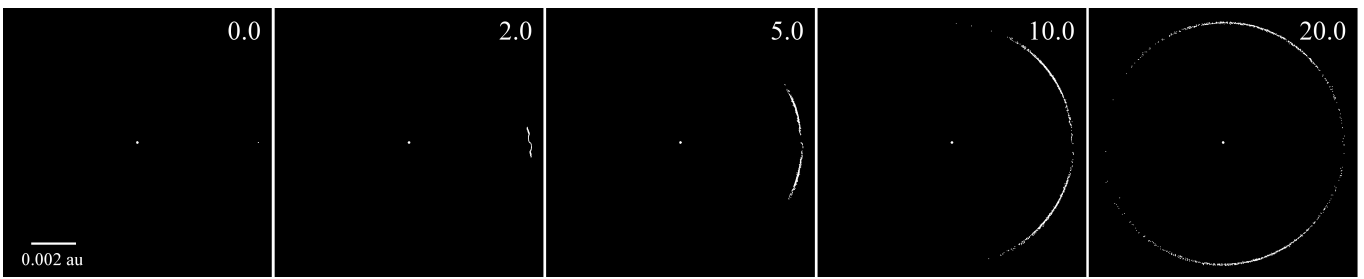}
}
\caption{
Spreading of stripped particles around the white dwarf 
(located at the centre) for the
mantle disruption in Fig. \ref{mandis} (simulation {\tt RandDiff19}). The
disrupting rubble pile is at the same position in each panel, centre-right.
Rubble pile particles have been inflated to enhance their visibility.  An animation
accompanying this figure is available online.
}
\label{ringspread}
\end{figure*}
%%%%%%%%%%%%%%%%% Figure Ecc

%%%%%%%%%%%%%%%%% Figure Ecc
\begin{figure*}
\centerline{
\includegraphics[width=18.0cm]{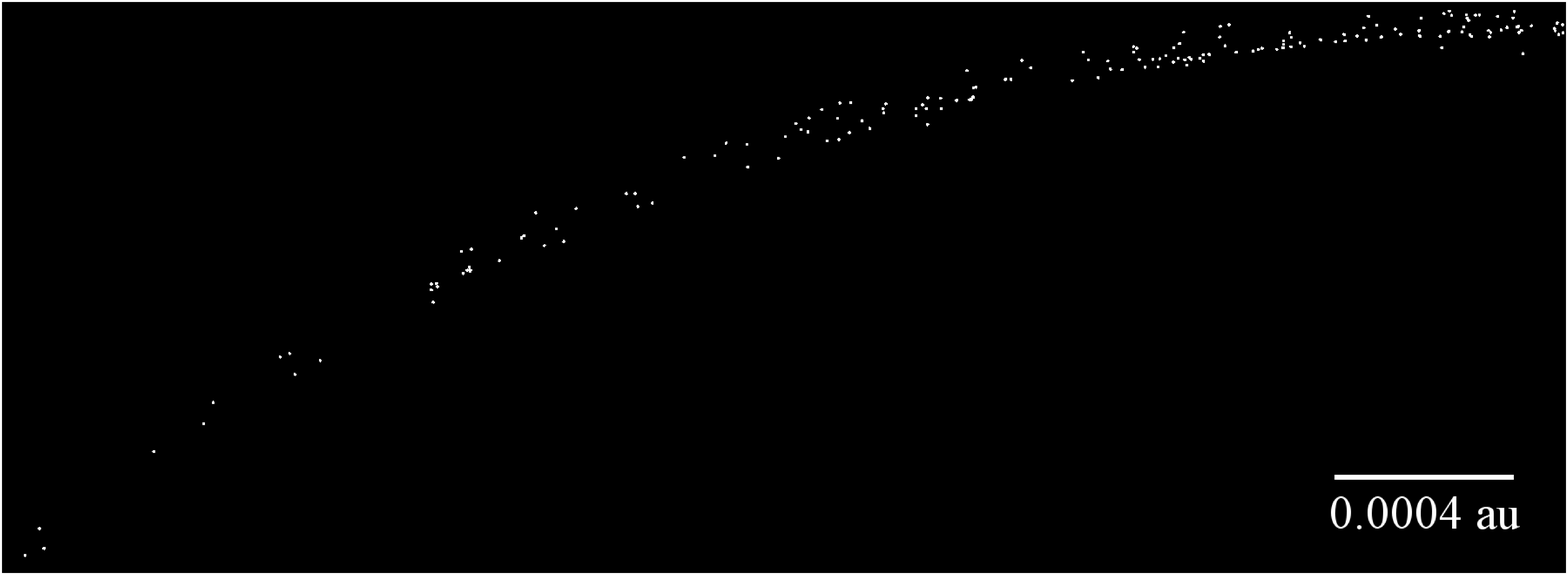}
}
\caption{
Close-up of the rightmost panel of Fig. \ref{ringspread} (simulation {\tt RandDiff19}), 
illustrating annular extent, clumpiness and voids.
Particles have been inflated to enhance their visibility.
}
\label{snaparc}
\end{figure*}
%%%%%%%%%%%%%%%%% Figure Ecc

\subsection{Simulation results for differentiated rubble piles}

Our results with homogeneous rubble piles could not explain the constant 
periodic signal or the transient signals over a two-year period.
Therefore, motivated by \cite{leietal2012}, who suggested that a differentiated
body might allow for partial disruption, we also adopted a differentiated rubble pile.
This assumption is realistic because the primary body could easily be large 
enough (like the asteroid Vesta; \citealt*{guretal2016,rapetal2016}) to be differentiated. 
For differentiated bodies, the packing type for rubble piles has a more pronounced
effect than in the homogeneous case. Therefore, for greater realism all of our 
differentiated rubble piles were randomly packed (consequently,
because of the lower packing efficiency, these rubble piles will disrupt more easily;
see Figs. \ref{rubpile} and \ref{ternden}).

\subsubsection{Physics of disruption}

The disruption of differentiated rubble piles is more complex than
those of homogeneous ones, and has been thoroughly described in \cite{canup2010}
and \cite{leietal2012}. We do not repeat their detailed analyses here, but
rather just emphasize some of the important points below, and focus instead
on the implications for the WD 1145+017 system. Overall, our 
simulations here are consistent with the behaviour seen in those works.

We characterize the outcome of disrupting a rubble pile with a mantle
and core in one of three ways: (i) no disruption, (ii) mantle disruption,
and (iii) full disruption. 

Mantle disruption is shown in Fig. \ref{mandis} (simulation {\tt RandDiff19})
and in the accompanying online animation. In this case, the green core
particles remain in place and hidden from view as the white mantle particles
are slowly stripped off. The streaming occurs at the L1 and L2 Lagrange points
after the rubble pile has been distorted into the shape of a lemon. This process
reproduces the schematic in Fig. A1 of \cite{rapetal2016}, except for
the major difference that in our numerical simulations, particles stream off from
both L1 and L2, as opposed to just L1. The streaming is not symmetrical from
both ends of the parent body, and up to 20 per cent more of the shorn-off particles
emanates from one Lagrangian point than the other (see Section \ref{disc}).
The streaming of particles increases the 
density of the rubble pile, which
allows it to subsequently resist disruption. Therefore, after about four orbits,
the mantle stripping becomes intermittent. The core density is high enough for
the core particles to remain protected from escape.

Full disruption occurs when, subsequent to mantle stripping, the remnant
core is not dense enough to resist breakup. This situation is visualized
in Fig. \ref{fulldis} (simulation {\tt RandDiff32}), whose rubble pile is equivalent
to that in Fig. \ref{mandis}, except placed on an $e=0.1$ orbit. After 
three orbits, most of the mantle has already separated and is in the process
of forming a ring.  After four orbits, the entire pile has catastrophically disrupted.

The trajectory of the stripped off particles forms a ring, just as in the homogeneous
case. In Fig. \ref{ringspread}, we illustrate this time sequence 
for the mantle disruption in Fig. \ref{mandis}. After about 10 orbits, the particles
have covered an arc halfway around the white dwarf. After about 20 orbits, a
full ring has formed, albeit one which contains inhomogeneities. We discuss ring
filling times in Section \ref{disc}.

Because the particles are stripped off from L1 and L2, they orbit at slightly different
distances than does the centre of the rubble pile. Consequently, these particles have slightly
different (both larger and shorter) orbital periods than the parent body; see Section \ref{disc}
for more details.

In order to visualize this difference in orbital period from our simulations,
in Fig. \ref{snaparc} we have zoomed-in on the top-left arc of the rightmost panel of Fig. 
\ref{ringspread}. This close-up illustrates both the scale of the annulus, and regions
of clumpiness and voids. For $M=10^{20}$~kg and $10^{22}$~kg parent bodies, the difference in
orbital periods from particles on each end of the annulus is on the order of, respectively, 
a couple tens of seconds, and about one hundred seconds. In this regard, the $M=10^{20}$~kg
case better matches the orbital period differences given by Fig. A3 and equation A11
of \cite{rapetal2016} and Table 1 of \cite{gaeetal2016}. Further, this mass is consistent
with the estimates given by \cite{guretal2016} and \cite{rapetal2016}. The slight excess 
difference that we see over
Rappaport et al.'s (2016) calculation is likely due to collisions in the forming ring.

\subsubsection{Transit model}

How do these disruption simulations relate to observable photometric transits?
WD~1145+017 features some of the most spectacular transit curves of any exoplanetary
system, with transit depths reaching 60\%, transit features appearing
and disappearing on a nightly basis, and some appearing over multiple
nights.  The periods of the individual transits are stable to a few seconds 
(Table 1 of \citealt*{gaeetal2016} and Table 4 of \citealt*{rapetal2016})
even though they differ by up to tens of seconds.
In other words, the individual periods are seen to be more stable than their 
spread among different fragments. \cite{rapetal2016} suggested that the transit 
curves in WD 1145+017 may be a result of fragments
that break off from a single asteroid. Because we have found that mantle stripping is intermittent, this 
process could produce fragments that occasionally obscure
the light from the white dwarf and create detectable dips in photometric light curves. 

Our gravity-only simulations are too simplistic to reproduce the detail in these transit curves,
particularly because they arise from dust- and gas-streaming off the fragments
rather than from geometrical blocking of the white dwarf by solid bodies. Nevertheless, we 
have created a transit model from our simulations by inflating the size of our particles 
which stream away from the disrupted rubble pile. The flux was calculated by dividing the face of 
the white dwarf into about 27000 pixels, and counting the number that were not obscured by the 
transiting fragments.

\begin{figure}
\centerline{
\includegraphics[width=8.7cm]{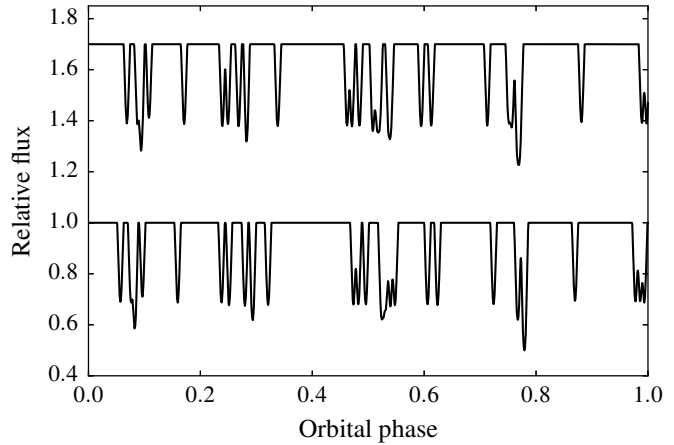}
}
\caption{
Illustrative model of photometric transit depths due to mantle stripping,
a process which intermittently streams particles off of a rubble pile.
Shown here are two consecutive orbits offset in flux.
The rubble pile used here was from simulation {\tt RandDiff19} (Fig. \ref{mandis}),
with particles suitably inflated. See the main text for more details.
}
\label{transit}
\end{figure}

Figure \ref{transit} displays the result, in two consecutive orbits offset by flux. Here, we have 
used only the particles that emanated from the rubble pile (that is shown in Fig. \ref{mandis}; 
simulation {\tt RandDiff19}) during the previous 50 orbits.  
The choice of 50 orbits is arbitrary, but represents the idea that fragments are only
``active'' -- i.e. expelling a cloud of gas or dust -- for a finite time\footnote{If we had not
implemented a cutoff, then the simulation would have been saturated as fragments were spread
into a ring, but not removed.}.
The time shown is a few hundred orbits after the start of the run, once the rate of
fragment escape has slowed.
Following \cite{gaeetal2016}, we also introduced two scaling factors: (i) the particle sizes 
were inflated to four times the 
size of the white dwarf (in order to achieve appropriate durations), and (ii) the transit depths 
were scaled such that the maximum depth 
was 0.5. The line-of-sight inclination was assumed to be offset by $2.25^{\circ}$ from an edge-on 
orientation, following the assumption in \cite{gaeetal2016}. 

The primary benefit of this simulation is to show that (i) the transit durations over a single
orbit are commensurate with those actually observed, and (ii) the non-uniformity of the transit
features may be reproduced by mantle stripping. Because the stripping is intermittent, this 
process may help explain the transience of some of the observed features.

%%%%%%%%%%%%%%%%% Figure Ecc
%%%%%%%%%%%%%%%%% Figure Ecc

\subsubsection{Density constraints}

Our simulations suggest that the process of mantle disruption 
might play an important role in the dynamics of WD 1145+017. A natural
accompanying question is, for what values of $\rho$ does mantle disruption occur?
In order to pursue an answer, and informed by the bounds imposed from
our homogenous rubble pile results, we have simulated differentiated
rubble piles with eleven different bulk core plus mantle 
densities from $2.5$ to $4.0$ g\,cm$^{-3}$.

\begin{figure*}
\centerline{
\includegraphics[width=8.7cm]{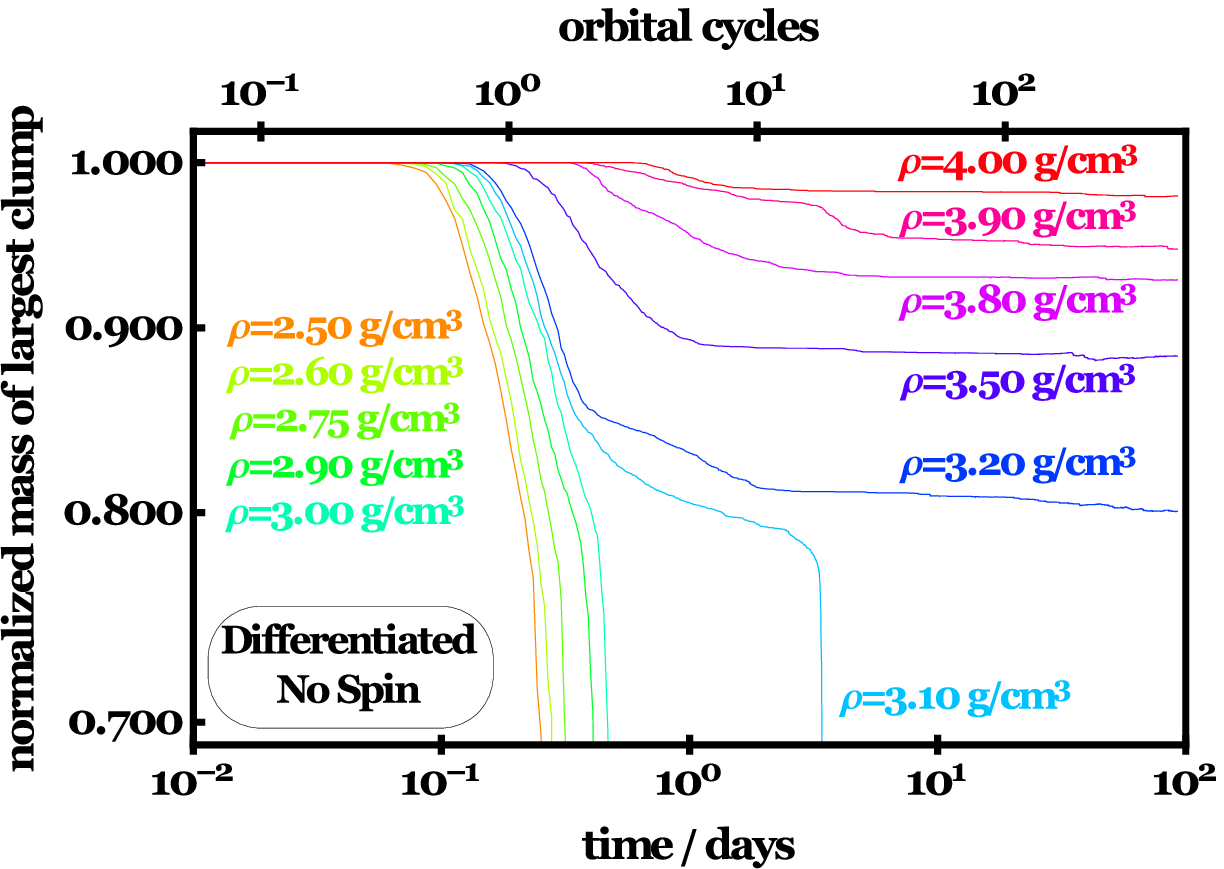}
\
\
\includegraphics[width=8.7cm]{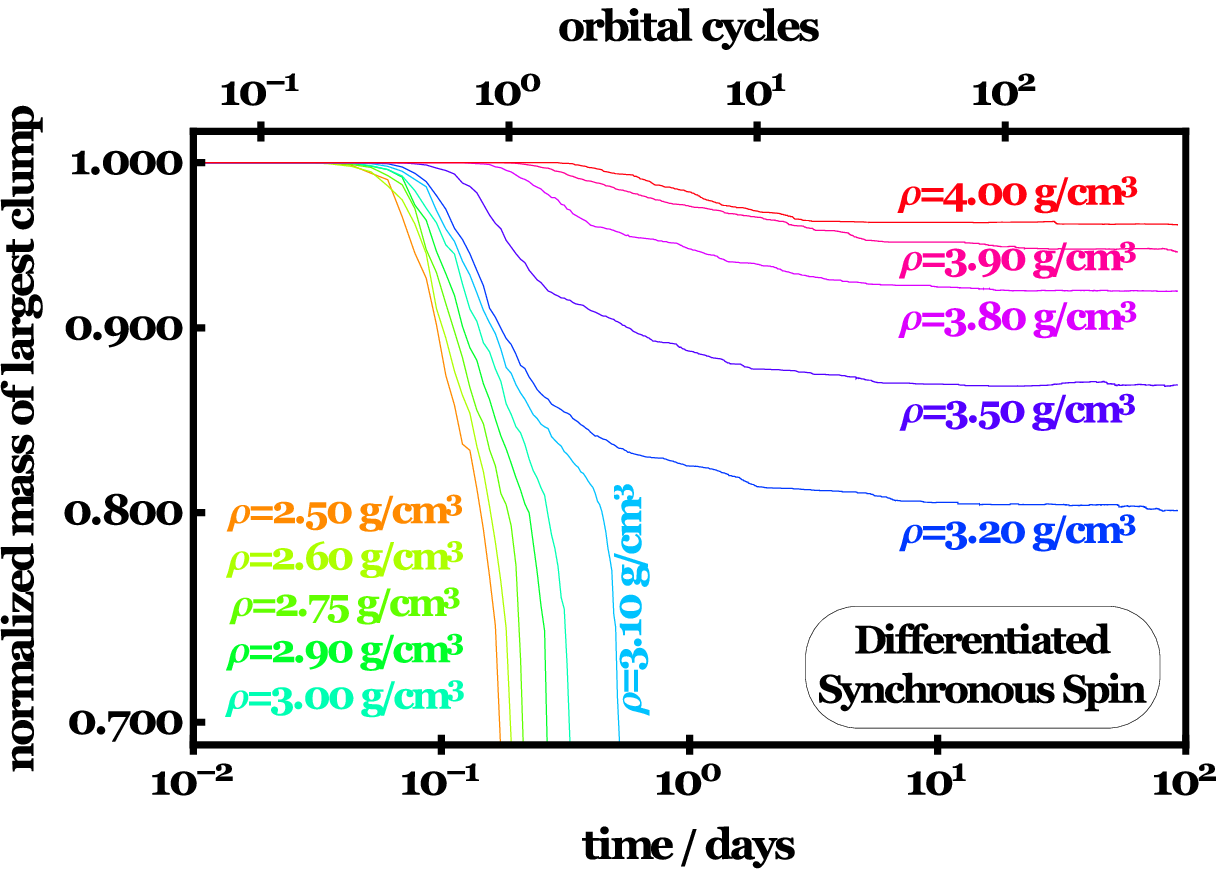}
}
\caption{
Disruption timescales for differentiated rubble piles, where the core
particle mass is quadruple that of the mantle particle mass (simulations {\tt RandDiff1} to {\tt RandDiff22}).  
The rubble piles
are on circular orbits and have either no spin (left panel) or synchronous spin, equivalent to being
tidally locked (right panel).
The slight decrease in mass at the tens of percent level seen in the curves with $\rho \ge 3.2$ g\,cm$^{-3}$
indicate mantle disruption. Full disruption occurs for $\rho \le 3.1$ g\,cm$^{-3}$.
This robust constraint on density, although specific to the rubble pile modelled, 
is similar to the robustness of the density constraints observed in Figs. \ref{heat} and \ref{Eccentricity}. 
}
\label{EccentricityDiff}
\end{figure*}

We present the results in Fig. \ref{EccentricityDiff}, which has a similar
format to Fig. \ref{Eccentricity} and contains the simulations labelled
{\tt RandDiff1} to {\tt RandDiff22} in Table \ref{simtable}. Here, however, mantle disruption is
indicated by a slight decrease, at the few to tens of per cent level, in 
normalized mass of the largest remaining clump, before levelling off.
Note that mantle or full disruption occurs in every simulation
in the figure. The amount of material lost decreases for higher
densities; at the high end ($\rho = 4.0$ g\,cm$^{-3}$),
in the left and right panels respectively, 2.1 and 3.9 per 
cent of the total mass was lost.  For $\rho < 3.2$ g\,cm$^{-3}$, 
mantle disruption can be observed to occur for a few hours before 
full disruption occurs more quickly.

Comparison of the two panels in the figure indicates that 
the initial spin of the rubble pile makes little difference to the
outcomes, except at the boundaries of the disruption regimes.
For the particular differentiated rubble piles we sampled in this
work (with a core four times more massive than the mantle) the transition
density between mantle disruption and full disruption occurs at  
$\rho = 3.1$ g\,cm$^{-3}$. At this density, a synchronously spinning
rubble pile disrupts an order of magnitude more quickly than
a rubble pile with no initial spin.

Although the boundary defined by $\rho = 3.1$ g\,cm$^{-3}$ is strongly
linked to the structure of the differentiated rubble pile that we
adopted, the clear division between disruption regimes on the
figure suggests that other rubble pile constructions will yield
similarly robust constraints.

%%%%%%%%%%%%%%%%% Figure Ecc
%%%%%%%%%%%%%%%%% Figure Ecc

\subsubsection{General constraints}

Other shape geometries may need to be considered when modelling disruption
of parent bodies in other systems. In fact, the results of our simulations might aid in future 
efforts, a prospect that we consider in this subsection.

\begin{figure}
\centerline{
\includegraphics[width=8.7cm]{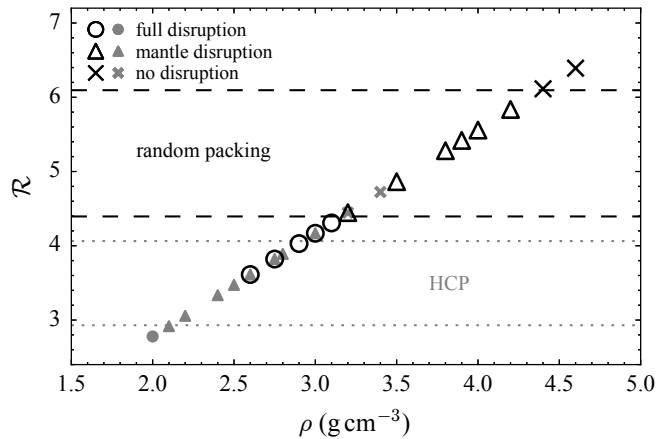}
}
\caption{
Boundaries between the regimes of ``full disruption'', ``mantle disruption'' and ``no disruption''.
The darker hollow symbols refer to simulations with randomly-packed
rubble piles, and the lighter filled symbols to outcomes with hexagonal-closely-packed (HCP) 
rubble piles.
The boundaries are approximated by the dashed lines, and are in good agreement with Fig. 2 of 
Leinhardt et al. (2012). 
}
\label{ternden}
\end{figure}

\begin{figure}
\centerline{
\includegraphics[width=8.7cm]{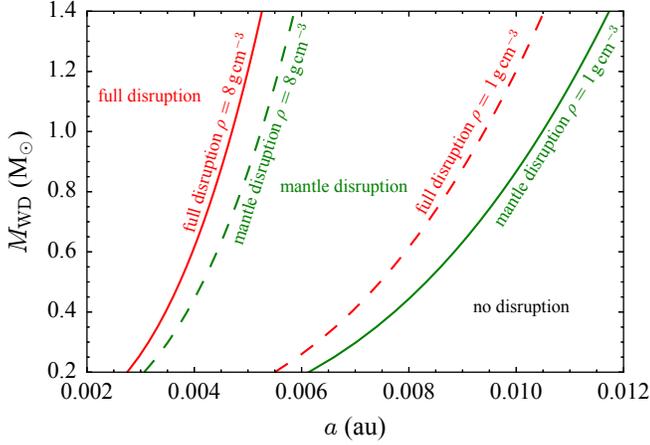}
}
\caption{
Pinpointing the three regimes where full disruption, mantle disruption
and no disruption occur as a function of stellar mass
($M_{\rm WD}$) and parent body semimajor axis ($a$). The regions are
identified by the horizontal labels. The curves
are based on the dashed lines in Fig. \ref{ternden}
and are applicable only for our randomly-packed rubble pile simulations.
}
\label{ternsemi}
\end{figure}

Unlike for homogeneous rubble piles, which either disrupt fully or not at all,
differentiated rubble piles could undergo no disruption, mantle disruption or
full disruption. In order to better quantify the parameter regimes
encompassing these ternary outcomes, we present Figs. \ref{ternden} and \ref{ternsemi}.

Figure \ref{ternden} links disruption with $\mathcal{R}$ and $\rho$ through
equations (\ref{zoec}) and (\ref{denper}). The darker symbols represent
simulations {\tt RandDiff2} through {\tt RandDiff11}, plus {\tt RandDiff23}, {\tt RandDiff24},
and {\tt RandDiff25}. The lighter symbols represent the outcomes for additional
simulations ({\tt HCPDiff1} to {\tt HCPDiff11}). The arbitrarily-drawn dashed
lines then approximate the critical values (4.4 and 6.1) of $\mathcal{R}$ which
separate the three regimes.

Figure \ref{ternden} can be compared directly to the bottom plot of Fig. 2 of
\cite{leietal2012}, which presents outcomes of tidal disruption simulations of
randomly-packed differentiated moons. Although they sampled a range
of $a$, and hence $P$, the agreement with the critical values of 4.4 and 6.1 is
good, to within one unit of $\mathcal{R}$ in each case. Any differences
could be attributed to the details of the packing geometry, including the bulk radius.

These critical values of 4.4 and 6.1 were then used to create Fig. \ref{ternsemi}, which
ignores our knowledge of $P$. This figure approximates the boundaries between the
disruption regimes (red = full disruption, green = mantle disruption, 
black = no disruption) in $(a,M_{\rm WD})$ space, for all possible white dwarf masses,
where two extreme densities (1~g\,cm$^{-3}$ and 8~g\,cm$^{-3}$) are given.
The solid lines provide absolute bounds. In the region to the left of the red solid
line, full disruption always occurs. In the region to the right of the green solid
line, no disruption ever occurs. In between those two lines, any three possibilities
may occur depending on the choice of $\rho$.
If the asteroid is less differentiated or centrally concentrated, then these curves would
all shift rightward.

Both figures represent useful tools for quick identification of a disruption regime for
future discoveries of disintegrating bodies around stars. These stars need not be
restricted to white dwarfs. In fact,
several bodies disintegrating around main sequence stars are now known 
\citep{rapetal2012,rapetal2014,croetal2014,bocetal2015,sanetal2015}.

%%%%%%%%%%%%%%%%% Figure Ecc
%%%%%%%%%%%%%%%%% Figure Ecc

%%%%%%%%%%%%%%%%% Figure Ecc
%%%%%%%%%%%%%%%%% Figure Ecc

%%%%%%%%%%%%%%%%% Figure Spread
\begin{figure}
\centerline{
\includegraphics[width=8.7cm]{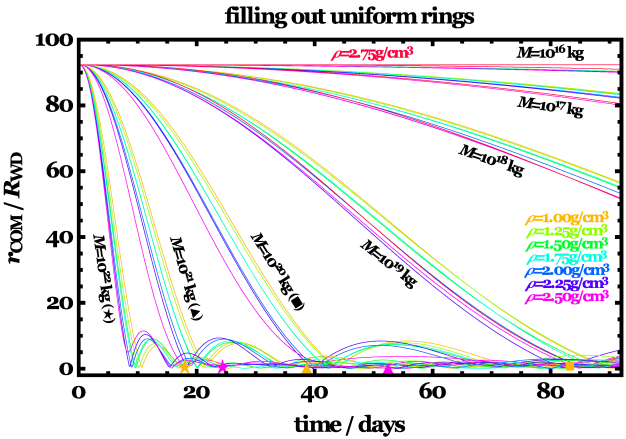}
}
\caption{
Ring filling times. Plotted is the time evolution of $r_{\rm COM}$,
which is the distance between the centre of mass of
the rubble pile particles and the centre of the white dwarf. 
As the rubble pile disrupts and a ring fills out, this distance decreases. 
When $r_{\rm COM} = 0$, a uniform ring has been created.
Inhomogeneities (clumpiness and voids) in the ring result in the ``bounces'' along 
the $x$-axis. Plotted symbols (stars for $M=10^{22}$ kg, triangles for $M=10^{21}$ kg, 
squares for $M=10^{20}$ kg; yellow for $\rho=1.0$~g\,cm$^{-3}$ and pink for $\rho=2.5$~g\,cm$^{-3}$) 
indicate the analytical predictions for the filling times 
of asteroids which are assumed to instantaneously disrupt at their orbital pericentres 
(equations \ref{peri} and \ref{filltime}).  
}
\label{spread}
\end{figure}
%%%%%%%%%%%%%%%%% Figure Spread

\section{Discussion}  \label{disc}

\subsection{Roche radius location}

We can now return to the question of whether the disrupting asteroid
is within the Roche radius of WD 1145+017. Consider that (1) the observational
data suggests that the asteroid has not undergone full disruption for over two years,
(2) we have shown that an asteroid which remains intact for this long has
$\rho \gtrsim 3.1$~g\,cm$^{-3}$ on a near-circular orbit, (3) the density which
corresponds to the Roche radius for this white dwarf is, for a solid,
tidally-locked body, $1.6$ g cm$^{-3}$,
(4) we have shown that mantle disruption can qualitatively reproduce
the intermittent transit features that are observed.

These statements imply that the parent body is {\it not} within the Roche radius
for its bulk density,
but rather just outside, and undergoing mantle disruption. Some caveats
which might negate this conclusion are if the asteroid's shape is
significantly non-spherical, the asteroid is differentiated in a complex
manner, or if the asteroid's mass contains a significant amount of
non-solid matter. These cases all represent viable, interesting and important
topics for future studies. Objects like 67P/Churyumov-Gerasimenko hint at the complexity
of small Solar system bodies, and similar bodies could maintain an internal
reserviour of volatiles even throughout the giant branch phases of stellar
evolution \citep{jurxu2010,juretal2012,malper2016}.

The details of the disruption which we have shown here are different
than those envisaged by \cite{rapetal2016}. Here, we see mass streaming
from both L1 and L2 points. Further, we find that the relative fraction
of particles escaping from L1 is slightly greater than those from L2,
with the disparity increasing for higher bulk densities. In particular,
for $\rho=3.9$~g\,cm$^{-3}$, 56 per cent come from L1, whereas for both
$\rho=4.0$~g\,cm$^{-3}$ and $\rho=4.2$~g\,cm$^{-3}$, 59 per cent originate from
L1. These differences
are observationally constrained, although only material with shorter orbits is visible
\citep{gaeetal2016,rapetal2016}. Perhaps the fragments with longer periods
are not currently active, or that the imbalance between streaming from L1 and L2
becomes larger with time.
{\it K2} data did show some very weak signals at longer periods, but their reality could not be
confirmed independently, as those signals are below the sensitivity threshold 
for ground-based observations. 

If the asteroid indeed lies just outside of the Roche radius,
then how did it arrive at a nearly-circular orbit at that location?
One possibility is that sustained gas ejection may be strong enough to appreciably
decay the orbit of close minor planets \citep[e.g. Eqs. 32-33 of][]{perchi2013},
although the exact manner of the orbital evolution may be nontrivial
\citep{bouetal2012,veretal2013b,doskal2016a,doskal2016b}.
An alternative (A. Johansen, private comm) is that the asteroid represents a 
second-generation minor planet which grew out of smaller debris from a disrupted
planetesimal that accumulated outside of the Roche radius, 
an idea previously proposed for Solar system moons \cite[e.g.][]{cricha2012}.
\cite{chaetal2011} suggested that the 
inner mid-sized moons of Saturn (Mimas, Enceladus, Tethys, Dione, and Rhea) could 
all have been formed by a viscously spreading massive ring that was itself the result 
of a large disruptive impact.
Phobos and Deimos (Mars's moons) could have formed close enough to their parent planet
that they have since spiraled in close to Mars due to 
tidal decay of the orbit \citep{roscha2012}.

\subsection{Ring filling time}

We have shown that disrupting rubble piles form rings.
How efficiently does the debris spread around the white dwarf? 
We approach this question both analytically and with the numerical output.
A key caveat to both approaches is that gas drag is neglected, which may
play a significant role in the WD 1145+107 system\footnote{Gas is likely
produced from sublimation and/or collisions of solid particles. The most recent substantial attempt
to model this interaction between gas and dust \citep{metetal2012}
indicates a complexity which is beyond the scope of this work.}.
Also, the analytical approach ignores
collisions, which are treated in the \textsc{PKDGRAV} simulations.

\subsubsection{Analytic filling times}

Assume the breakup is instantaneous and occurs at a distance $r_{\rm b}$ and that the 
particles composing the parent body are collisionless.  Then, from Eq. (25) of 
\cite{veretal2014b}, a breakup event will fill out a complete ring in space in a 
filling time $t_{\rm fill}$ given by

\[
\frac{t_{\rm fill}}{P}
=
r_{\rm b}^{\frac{3}{2}}
\bigg[  
\bigg\lbrace 
\frac
{r_{\rm b}^2 + 2aR - r_{\rm b}R}
{r_{\rm b}-R}
\bigg\rbrace^{\frac{3}{2}}
\]

\begin{equation}
-
\bigg\lbrace 
\frac
{r_{\rm b}^2 - 2a\times{\rm min}(r_{\rm crit}-r_{\rm b}, R) + r_{\rm b}{\rm min}(r_{\rm crit}-r_{\rm b}, R)}
{r_{\rm b}+{\rm min}(r_{\rm crit}-r_{\rm b}, R)}
\bigg\rbrace^{\frac{3}{2}}
\bigg]^{-1}
,
\nonumber
\label{fill1}
\end{equation}

\noindent{}where

\begin{equation}
r_{\rm crit} = \frac{2ar_{\rm b}}{\left(1 + \frac{M}{M_{\rm WD}} \right) \left(2a - r_{\rm b}\right)}
\approx
\frac{2ar_{\rm b}}{2a - r_{\rm b}}
\end{equation} 

\noindent{}is the distance at which a particle's orbit becomes parabolic (will escape from the system). For the parent 
bodies orbiting WD 1145+017, if we suppose they break up at pericentre, then $r_{\rm b} = a(1-e)$ and 

\

\[
\frac{t_{\rm fill}}{P} = 
\left(1 - e\right)^{\frac{3}{2}}
\]

\[
\times\left[
\left(
\frac{a\left(1-e\right)^2 +R\left(1+e\right)}
     {a\left(1-e\right)   -R}
\right)^{\frac{3}{2}}
-
\left(
\frac{a\left(1-e\right)^2 -R\left(1+e\right)}
     {a\left(1-e\right)   +R}
\right)^{\frac{3}{2}}
\right]^{-1}
\]
\vspace{-0.9em}
\begin{equation}
\ \ \ \ \ \approx 
\frac{a}{6R}
\left(1-e\right)^2
\label{peri}
\end{equation}
\vspace{-0.9em}
\begin{equation}
\ \ \ \ \ = 1.3 \times 10^3 
\left(\frac{a}{0.00535 \ {\rm au}}\right)
\left(\frac{R}{100 \ {\rm km}}\right)^{-1}
\left(1-e\right)^2
\end{equation}

\noindent{}corresponding to a filling time of

\begin{equation}
t_{\rm fill} \approx 250 \ {\rm days} 
\left(\frac{a}{0.00535 \ {\rm au}}\right)
\left(\frac{R}{100 \ {\rm km}}\right)^{-1}
\left(1-e\right)^2
.
\label{filltime}
\end{equation}

\noindent{}Note that the filling time is dependent on radius, and hence mass, for a given density (unlike for disruption). The reason is because the physical separation of the L1 and L2 points depends on the parent body size. Hence, particles leaving from the Lagrangian points will have larger initial orbital period differences for larger parent body sizes.

\subsubsection{Numerical filling times}

Numerically, one way to determine how quickly particles spread into a ring
is to consider the time evolution of the centre of mass of the particles,
$r_{\rm COM}$. Initially, $r_{\rm COM} \approx a \approx 92 R_{\rm WD}$, assuming
$R_{\rm WD} = 8750$ km (a fiducial white dwarf radius; \citealt*{veretal2014b}). 
As the rubble pile disrupts and the particles spread
into a ring, this centre of mass will gradually move towards the centre of
the white dwarf. Consequently, a uniform ring is formed as $r_{\rm COM} \rightarrow 0$.
However, as $r_{\rm COM}$ approaches zero, it oscillates as ring particles 
overtake each other, and potentially collide. 
This movement ceases to be monotonic, and inhomogeneities in the ring will show 
up as nonzero values of $r_{\rm COM}$.

\subsubsection{Results}

In Fig. \ref{spread}, we plot the time evolution of $r_{\rm COM}$
for all homogeneous HCP rubble piles that disrupted. Overplotted
as stars, triangles and a square are the analytical estimates of the
filling time from equation (\ref{filltime}).
Quantifying the extent of the agreement between the analytics and
numerics is not possible
unless one defines the meaning of a ring that has been ``filled'':
the curves in Fig. \ref{spread} appear to ``bounce'' on the $x$-axis several
times before settling. Equation (\ref{filltime}) best reproduces a point
in-between the second and third bounce.

Some trends from the figure are worth noting, particularly because
of their potential use for interpreting future observations: (i) the circularization
time decreases with increasing mass, (ii) the ``bounciness'' increases
with increasing mass, (iii) rubble piles with $M \gtrsim 10^{20}$ kg generally
spread out into full rings within about three months, and (iv) for a given mass, higher
density rubble piles more quickly achieve particle coverage throughout the orbit, but
do not necessarily evenly fill out the ring more quickly or smoothly.

The figure demonstrates that the ring filling times range from 10 to 100 days
for $10^{19}$~kg~$\lesssim M \lesssim$~$10^{22}$~kg. Although this timescale
fits within the baseline of observations, the ring is not directly observed.
One possible reason is that the ring is collisionally eroded; another is that
the newly disrupted pieces of mantle are not active. However, theoretical models which
include dust and/or gas might better link observed infrared excess (indicating
dust) or circumstellar gas \citep{xuetal2016} with ring formation.

\section{Conclusion}

The properties of the asteroid disintegrating around white dwarf WD 1145+017
are poorly constrained observationally. Theoretical work, however, can help 
remedy this shortfall, and in this case suggests that the disintegrating
asteroid orbiting WD 1145+017 appears to reside just outside of the bulk
density Roche radius.

In particular, we have modelled the tidal disruption of strengthless
rubble piles with an orbital period equal to that of the longest-lasting
observed transit signature. We found robust
constraints on density (Figs. \ref{heat} and \ref{EccentricityDiff})
and eccentricity (Fig. \ref{Eccentricity}), and weak constraints on mass
(Fig. \ref{heat}) and spin (Fig. \ref{EccentricityDiff}) (but see the last paragraph
of Section 5.2.1). By modeling both homogeneous
and differentiated rubble piles, we found that $\rho \lesssim 2.75$\,g~cm$^{-3}$
ensures disruption within one day, whereas $\rho \gtrsim 3.10$\,g~cm$^{-3}$ guarantees
that rubble piles on circular orbits will remain intact for at least two years. Nevertheless, 
the intact differentiated
rubble piles all undergo mantle disruption, which produces intermittent streams of
particles which may contribute to the observed photometric transit dips (Fig. \ref{transit}). 
If the eccentricity of the disrupting object exceeds 0.130,
then it is unlikely to remain intact for more than one month unless
$\rho > 4.0$ g~cm$^{-3}$.

Useful ancillary results include figures which may be applicable for studies of
disruption around other stars. These figures include a mass-free relation between orbital period
and density (Fig. \ref{dengrid}, and equation \ref{denper}) and parameter-space 
locations where we can expect to find mantle disruption versus full disruption 
(Figs. \ref{ternden} and \ref{ternsemi}).

We must caution that our seemingly robust results for WD 1145+017 rely 
on several assumptions: 
(i) the parent body is spherical, 
(ii) the parent body is strengthless, and
(iii) the parent body or its fragments are not affected
by the extant dust or gas in the system.  Relaxing these assumptions
\citep[e.g.][]{metetal2012,movetal2012,schetal2012}  
as new observations warrant might 
place stricter physical constraints on the system.

\section*{Acknowledgements}

We thank the referee for a careful read-through of the manuscript, and valuable comments. DV and BTG have received funding from the European Research Council under the European Union's Seventh Framework Programme (FP/2007-2013)/ERC Grant Agreement n. 320964 (WDTracer).  PJC and ZML acknowledge support from the Natural Environment Research Council (grant number: NE/K004778/1).

\appendix

\section{Simulation table}

\onecolumn

\begin{center}
\begin{longtable}{c c c c c c c c c c c}
\caption{Summary of simulations. The radii are rounded to two significant digits.
Emboldended entries indicates important variables that were varied within each set.}
\label{simtable}
\\
\hline 
\multicolumn{1}{c}{simulation} & \multicolumn{1}{c}{packing}      & \multicolumn{1}{c}{differe-} &
\multicolumn{1}{c}{number of}  & \multicolumn{1}{c}{density}      & \multicolumn{1}{c}{mass}           &
\multicolumn{1}{c}{radius}     & \multicolumn{1}{c}{$e$} & \multicolumn{1}{c}{spin}           &
\multicolumn{1}{c}{duration}   & \multicolumn{1}{c}{outcome}
\\ 
\multicolumn{1}{c}{name}       & \multicolumn{1}{c}{type}         & \multicolumn{1}{c}{-ntiated}              &
\multicolumn{1}{c}{particles}  & \multicolumn{1}{c}{(g cm$^{-3}$)}   & \multicolumn{1}{c}{(kg)}           &
\multicolumn{1}{c}{(km)}       & \multicolumn{1}{c}{}            & \multicolumn{1}{c}{ }              &
\multicolumn{1}{c}{(days)}     & \multicolumn{1}{c}{(disruption type)}
\\
\multicolumn{1}{c}{}           & \multicolumn{1}{c}{}        & \multicolumn{1}{c}{}           &
\multicolumn{1}{c}{}           & \multicolumn{1}{c}{}        & \multicolumn{1}{c}{}           &
\multicolumn{1}{c}{}           & \multicolumn{1}{c}{}        & \multicolumn{1}{c}{}           &
\multicolumn{1}{c}{}           & \multicolumn{1}{c}{}
\endfirsthead

\multicolumn{11}{c}%
{{\bfseries \tablename\ \thetable{} -- continued from previous page}} \\
\hline 
\multicolumn{1}{c}{simulation} & \multicolumn{1}{c}{packing}      & \multicolumn{1}{c}{differe-}  &
\multicolumn{1}{c}{number of}  & \multicolumn{1}{c}{density}      & \multicolumn{1}{c}{mass}            &
\multicolumn{1}{c}{radius}     & \multicolumn{1}{c}{$e$} & \multicolumn{1}{c}{spin}            &
\multicolumn{1}{c}{duration}   & \multicolumn{1}{c}{outcome}
\\ 
\multicolumn{1}{c}{name}       & \multicolumn{1}{c}{type}         & \multicolumn{1}{c}{-ntiated}               & 
\multicolumn{1}{c}{particles}  & \multicolumn{1}{c}{(g cm$^{-3}$)}   & \multicolumn{1}{c}{(kg)}            &
\multicolumn{1}{c}{(km)}       & \multicolumn{1}{c}{}            & \multicolumn{1}{c}{ }               &
\multicolumn{1}{c}{(days)}     & \multicolumn{1}{c}{(disruption type)}
\\
\multicolumn{1}{c}{}           & \multicolumn{1}{c}{}        & \multicolumn{1}{c}{}           &
\multicolumn{1}{c}{}           & \multicolumn{1}{c}{}        & \multicolumn{1}{c}{}           &
\multicolumn{1}{c}{}           & \multicolumn{1}{c}{}        & \multicolumn{1}{c}{}           &
\multicolumn{1}{c}{}           & \multicolumn{1}{c}{}
\endhead

\hline \multicolumn{7}{r}{{Continued on next page}} \\ \hline
\endfoot

\hline \hline
\endlastfoot

%\hline
%simulation &  packing  & differentiated         & number of  & density      & mass      & radius     & eccentricity & spin & duration & outcome \\
%   name    &   type    &    & particles  & (g cm$^{-3}$) & (kg)      &  (km)      &              &      &  (days)  & (disruption type) \\
\hline
HCP1   &  hexagonal  &     no   &  5003  & 1.00      &  ${\mathbf{10^{16}}}$  & $14$   & 0.00          & 0 & 90  &  full  \\
HCP2   &  hexagonal  &     no   &  5003   & 1.00      & ${\mathbf{10^{17}}}$  & $29$  & 0.00          & 0 & 90  &  full \\
HCP3   &  hexagonal  &     no   &  5003   & 1.00      & ${\mathbf{10^{18}}}$  & $62$  & 0.00          & 0 & 90  &  full \\
HCP4   &  hexagonal  &     no   &  5003   & 1.00      & ${\mathbf{10^{19}}}$  & $130$  & 0.00          & 0 & 90  &  full \\
HCP5   &  hexagonal  &     no   &  5003   & 1.00      & ${\mathbf{10^{20}}}$  & $290$  & 0.00          & 0 & 90  &  full \\
HCP6   &  hexagonal  &     no   &  5003   & 1.00      & ${\mathbf{10^{21}}}$  & $620$  & 0.00          & 0 & 90  &  full \\
HCP7   &  hexagonal  &     no   &  5003   & 1.00      & ${\mathbf{10^{22}}}$  & $1300$  & 0.00          & 0 & 90  &  full \\
\hline
HCP8   &  hexagonal  &     no   &  5003   & 1.25      & ${\mathbf{10^{16}}}$  & $12$   & 0.00          & 0 & 90  &  full \\
HCP9   &  hexagonal  &     no   &  5003   & 1.25      & ${\mathbf{10^{17}}}$  & $27$  & 0.00          & 0 & 90  &  full \\
HCP10  &  hexagonal  &     no   &  5003   & 1.25      & ${\mathbf{10^{18}}}$  & $58$  & 0.00          & 0 & 90  &  full \\
HCP11  &  hexagonal  &     no   &  5003   & 1.25      & ${\mathbf{10^{19}}}$  & $120$  & 0.00          & 0 & 90  &  full \\
HCP12  &  hexagonal  &     no   &  5003   & 1.25      & ${\mathbf{10^{20}}}$  & $270$  & 0.00          & 0 & 90  &  full \\
HCP13  &  hexagonal  &     no   &  5003   & 1.25      & ${\mathbf{10^{21}}}$  & $580$  & 0.00          & 0  & 90  &  full\\
HCP14  &  hexagonal  &     no   &  5003   & 1.25      & ${\mathbf{10^{22}}}$  & $1200$  & 0.00          & 0  & 90  &  full\\
\hline
HCP15   &  hexagonal  &     no  &  5003    & 1.50      & ${\mathbf{10^{16}}}$  & $12$   & 0.00          & 0 & 90  &  full \\
HCP16   &  hexagonal  &     no  &  5003    & 1.50      & ${\mathbf{10^{17}}}$  & $25$  & 0.00          & 0 & 90  &  full \\
HCP17  &  hexagonal  &     no   &  5003   & 1.50      & ${\mathbf{10^{18}}}$  & $54$  & 0.00          & 0 & 90  &  full \\
HCP18  &  hexagonal  &     no   &  5003   & 1.50      & ${\mathbf{10^{19}}}$  & $120$  & 0.00          & 0 & 90  &  full \\
HCP19  &  hexagonal  &     no   &  5003   & 1.50      & ${\mathbf{10^{20}}}$  & $250$  & 0.00          & 0 & 90  &  full \\
HCP20  &  hexagonal  &     no   &  5003   & 1.50      & ${\mathbf{10^{21}}}$  & $540$  & 0.00          & 0  & 90  &  full\\
HCP21  &  hexagonal  &     no   &  5003   & 1.50      & ${\mathbf{10^{22}}}$  & $1200$  & 0.00          & 0  & 90  &  full\\
\hline
HCP22   &  hexagonal  &     no  &  5003    & 1.75      & ${\mathbf{10^{16}}}$  & $11$   & 0.00          & 0 & 90  &  full \\
HCP23   &  hexagonal  &     no  &  5003    & 1.75      & ${\mathbf{10^{17}}}$  & $24$  & 0.00          & 0 & 90  &  full \\
HCP24  &  hexagonal  &     no   &  5003   & 1.75      & ${\mathbf{10^{18}}}$  & $51$  & 0.00          & 0 & 90  &  full \\
HCP25  &  hexagonal  &     no   &  5003   & 1.75      & ${\mathbf{10^{19}}}$  & $110$  & 0.00          & 0 & 90  &  full \\
HCP26  &  hexagonal  &     no   &  5003   & 1.75      & ${\mathbf{10^{20}}}$  & $240$  & 0.00          & 0 & 90  &  full \\
HCP27  &  hexagonal  &     no   &  5003   & 1.75      & ${\mathbf{10^{21}}}$  & $510$  & 0.00          & 0  & 90  &  full\\
HCP28  &  hexagonal  &     no   &  5003   & 1.75      & ${\mathbf{10^{22}}}$  & $1100$  & 0.00          & 0  & 90  &  full\\
\hline
HCP29   &  hexagonal  &     no  &  5003    & 2.00      & ${\mathbf{10^{16}}}$  & $11$   & 0.00          & 0 & 90  &  full \\
HCP30   &  hexagonal  &     no  &  5003    & 2.00      & ${\mathbf{10^{17}}}$  & $23$  & 0.00          & 0 & 90  &  full \\
HCP31  &  hexagonal  &     no   &  5003   & 2.00      & ${\mathbf{10^{18}}}$  & $49$  & 0.00          & 0 & 90  &  full \\
HCP32  &  hexagonal  &     no   &  5003   & 2.00      & ${\mathbf{10^{19}}}$  & $110$  & 0.00          & 0 & 90  &  full \\
HCP33  &  hexagonal  &     no   &  5003   & 2.00      & ${\mathbf{10^{20}}}$  & $230$  & 0.00          & 0 & 90  &  full \\
HCP34  &  hexagonal  &     no   &  5003   & 2.00      & ${\mathbf{10^{21}}}$  & $490$  & 0.00          & 0  & 90  &  full\\
HCP35  &  hexagonal  &     no   &  5003   & 2.00      & ${\mathbf{10^{22}}}$  & $1100$  & 0.00          & 0  & 90  &  full\\
\hline
HCP36   &  hexagonal  &     no   &  5003   & 2.25      & ${\mathbf{10^{16}}}$  & $10$   & 0.00          & 0 & 90  &  full  \\
HCP37   &  hexagonal  &     no   &  5003   & 2.25      & ${\mathbf{10^{17}}}$  & $22$  & 0.00          & 0 & 90  &  full \\
HCP38   &  hexagonal  &     no   &  5003   & 2.25      & ${\mathbf{10^{18}}}$  & $47$  & 0.00          & 0 & 90  &  full \\
HCP39   &  hexagonal  &     no   &  5003   & 2.25      & ${\mathbf{10^{19}}}$  & $102$  & 0.00          & 0 & 90  &  full \\
HCP40   &  hexagonal  &     no   &  5003   & 2.25      & ${\mathbf{10^{20}}}$  & $220$  & 0.00          & 0 & 90  &  full \\
HCP41   &  hexagonal  &     no   &  5003   & 2.25      & ${\mathbf{10^{21}}}$  & $470$  & 0.00          & 0 & 90  &  full \\
HCP42   &  hexagonal  &     no   &  5003   & 2.25      & ${\mathbf{10^{22}}}$  & $1000$  & 0.00          & 0 & 90  &  full \\
\hline
HCP43   &  hexagonal  &     no  & 5003   & 2.50      & ${\mathbf{10^{16}}}$  & $9.9$   & 0.00          & 0 & 90  &  full  \\
HCP44   &  hexagonal  &     no  & 5003   & 2.50      & ${\mathbf{10^{17}}}$  & $21$  & 0.00          & 0 & 90  &  full \\
HCP45   &  hexagonal  &     no  & 5003   & 2.50      & ${\mathbf{10^{18}}}$  & $46$  & 0.00          & 0 & 90  &  full \\
HCP46   &  hexagonal  &     no  & 5003   & 2.50      & ${\mathbf{10^{19}}}$  & $98$  & 0.00          & 0 & 90  &  full \\
HCP47   &  hexagonal  &     no  & 5003   & 2.50      & ${\mathbf{10^{20}}}$  & $210$  & 0.00          & 0 & 90  &  full \\
HCP48   &  hexagonal  &     no  & 5003   & 2.50      & ${\mathbf{10^{21}}}$  & $460$  & 0.00          & 0 & 90  &  full \\
HCP49   &  hexagonal  &     no  & 5003   & 2.50      & ${\mathbf{10^{22}}}$  & $990$  & 0.00          & 0 & 90  &  full \\
\hline
HCP50   &  hexagonal  &     no  & 5003   & 2.75      & ${\mathbf{10^{16}}}$  & $9.5$   & 0.00          & 0 & 90  &  full  \\
HCP51   &  hexagonal  &     no  & 5003   & 2.75      & ${\mathbf{10^{17}}}$  & $21$  & 0.00          & 0 & 90  &  full \\
HCP52   &  hexagonal  &     no  & 5003   & 2.75      & ${\mathbf{10^{18}}}$  & $44$  & 0.00          & 0 & 90  &  full \\
HCP53   &  hexagonal  &     no  & 5003    & 2.75      & ${\mathbf{10^{19}}}$  & $95$  & 0.00          & 0 & 90  &  full \\
HCP54   &  hexagonal  &     no  & 5003   & 2.75      & ${\mathbf{10^{20}}}$  & $210$  & 0.00          & 0 & 90  &  none \\
HCP55   &  hexagonal  &     no  & 5003   & 2.75      & ${\mathbf{10^{21}}}$  & $440$  & 0.00          & 0 & 90  &  none \\
HCP56   &  hexagonal  &     no  & 5003   & 2.75      & ${\mathbf{10^{22}}}$  & $950$  & 0.00          & 0 & 90  &  none \\
\hline
HCP57   &  hexagonal  &     no  & 5003   & 3.00      & $\mathbf{10^{16}}$  & $9.3$   & 0.00          & 0 & 90  &  none  \\
HCP58   &  hexagonal  &     no  & 5003   & 3.00      & $\mathbf{10^{17}}$  & $20$  & 0.00          & 0 & 90  &  none \\
HCP59   &  hexagonal  &     no  & 5003   & 3.00      & $\mathbf{10^{18}}$  & $43$  & 0.00          & 0 & 90  &  none \\
HCP60   &  hexagonal  &     no  & 5003   & 3.00      & $\mathbf{10^{19}}$  & $93$  & 0.00          & 0 & $\mathbf{180}$  &  none \\
HCP61   &  hexagonal  &     no  & $\mathbf{3985}$   & 3.00      & $\mathbf{10^{19}}$  & $93$  & 0.00          & 0 & $\mathbf{365}$  &  none \\
HCP62   &  hexagonal  &     no  & $\mathbf{3011}$   & 3.00      & $\mathbf{10^{19}}$  & $93$  & 0.00          & 0 & $\mathbf{730}$  &  none \\
HCP63   &  hexagonal  &     no  & 5003   & 3.00      & $\mathbf{10^{20}}$  & $200$  & 0.00          & 0 & 90  &  none \\
HCP64   &  hexagonal  &     no  & 5003   & 3.00      & $\mathbf{10^{21}}$  & $430$  & 0.00          & 0 & 90  &  none \\
HCP65   &  hexagonal  &     no  & 5003   & 3.00      & $\mathbf{10^{22}}$  & $930$  & 0.00          & 0 & $\mathbf{180}$  &  none \\
HCP66   &  hexagonal  &     no  & $\mathbf{3985}$   & 3.00      & $\mathbf{10^{22}}$  & $930$  & 0.00          & 0 & $\mathbf{365}$  &  none \\
HCP67   &  hexagonal  &     no  & $\mathbf{3011}$   & 3.00      & $\mathbf{10^{22}}$  & $930$  & 0.00          & 0 & $\mathbf{730}$  &  none \\
\hline
HCP68   &  hexagonal  &     no  & 5003   & 3.25      & $\mathbf{10^{16}}$  & $9.0$   & 0.00          & 0 & 90  &   none  \\
HCP69   &  hexagonal  &     no  & 5003   & 3.25      & $\mathbf{10^{17}}$  & $19$  & 0.00          & 0 & 90  &   none \\
HCP70   &  hexagonal  &     no  & 5003   & 3.25      & $\mathbf{10^{18}}$  & $42$  & 0.00          & 0 & 90  &   none \\
HCP71   &  hexagonal  &     no  & 5003   & 3.25      & $\mathbf{10^{19}}$  & $90$  & 0.00          & 0 & $\mathbf{180}$  &  none \\
HCP72   &  hexagonal  &     no  & $\mathbf{3985}$   & 3.25      & $\mathbf{10^{19}}$  & $90$  & 0.00          & 0 & $\mathbf{365}$  &  none \\
HCP73   &  hexagonal  &     no  & $\mathbf{3011}$   & 3.25      & $\mathbf{10^{19}}$  & $90$  & 0.00          & 0 & $\mathbf{730}$  &  none \\
HCP74   &  hexagonal  &     no  & 5003   & 3.25      & $\mathbf{10^{20}}$  & $190$  & 0.00          & 0 & 90  &  none \\
HCP75   &  hexagonal  &     no  & 5003   & 3.25      & $\mathbf{10^{21}}$  & $420$  & 0.00          & 0 & 90  &  none \\
HCP76   &  hexagonal  &     no  & 5003   & 3.25      & $\mathbf{10^{22}}$  & $900$  & 0.00          & 0 & $\mathbf{180}$  &  none \\
HCP77   &  hexagonal  &     no  & $\mathbf{3985}$   & 3.25      & $\mathbf{10^{22}}$  & $900$  & 0.00          & 0 & $\mathbf{365}$  &  none \\
HCP78   &  hexagonal  &     no  & $\mathbf{3011}$   & 3.25      & $\mathbf{10^{22}}$  & $900$  & 0.00          & 0 & $\mathbf{730}$  &  none \\
\hline
HCP79   &  hexagonal  &     no  & 5003   & 3.50      & $\mathbf{10^{16}}$  & $8.8$   & 0.00          & 0 & 90  &   none  \\
HCP80   &  hexagonal  &     no  & 5003   & 3.50      & $\mathbf{10^{17}}$  & $19$  & 0.00          & 0 & 90  &   none \\
HCP81   &  hexagonal  &     no  & 5003   & 3.50      & $\mathbf{10^{18}}$  & $41$  & 0.00          & 0 & 90  &   none \\
HCP82   &  hexagonal  &     no  & 5003   & 3.50      & $\mathbf{10^{19}}$  & $88$  & 0.00          & 0 & $\mathbf{180}$  &   none \\
HCP83   &  hexagonal  &     no  & $\mathbf{3985}$   & 3.50      & $\mathbf{10^{19}}$  & $88$  & 0.00          & 0 & $\mathbf{365}$  &  none \\
HCP84   &  hexagonal  &     no  & $\mathbf{3011}$   & 3.50      & $\mathbf{10^{19}}$  & $88$  & 0.00          & 0 & $\mathbf{730}$  &  none \\
HCP85   &  hexagonal  &     no  & 5003   & 3.50      & $\mathbf{10^{20}}$  & $190$  & 0.00          & 0 & 90  &  none \\
HCP86   &  hexagonal  &     no  & 5003   & 3.50      & $\mathbf{10^{21}}$  & $410$  & 0.00          & 0 & 90  &  none \\
HCP87   &  hexagonal  &     no  & 5003   & 3.50      & $\mathbf{10^{22}}$  & $880$  & 0.00          & 0 & $\mathbf{180}$  &  none \\
HCP88   &  hexagonal  &     no  & $\mathbf{3985}$   & 3.50      & $\mathbf{10^{22}}$  & $880$  & 0.00          & 0 & $\mathbf{365}$  &  none \\
HCP89   &  hexagonal  &     no  & $\mathbf{3011}$   & 3.50      & $\mathbf{10^{22}}$  & $880$  & 0.00          & 0 & $\mathbf{730}$  &  none \\
\hline
HCP90   &  hexagonal  &     no  & 5003   & 3.75      & $\mathbf{10^{16}}$  & $8.6$   & 0.00          & 0 & 90  &   none  \\
HCP91   &  hexagonal  &     no  & 5003   & 3.75      & $\mathbf{10^{17}}$  & $19$  & 0.00          & 0 & 90  &   none \\
HCP92   &  hexagonal  &     no  & 5003   & 3.75      & $\mathbf{10^{18}}$  & $40$  & 0.00          & 0 & 90  &   none \\
HCP93   &  hexagonal  &     no  & 5003   & 3.75      & $\mathbf{10^{19}}$  & $86$  & 0.00          & 0 & $\mathbf{180}$  &   none \\
HCP94   &  hexagonal  &     no  & $\mathbf{3985}$   & 3.75      & $\mathbf{10^{19}}$  & $86$  & 0.00          & 0 & $\mathbf{365}$  &  none \\
HCP95   &  hexagonal  &     no  & $\mathbf{3011}$   & 3.75      & $\mathbf{10^{19}}$  & $86$  & 0.00          & 0 & $\mathbf{730}$  &  none \\
HCP96   &  hexagonal  &     no  & 5003   & 3.75      & $\mathbf{10^{20}}$  & $190$  & 0.00          & 0 & 90  &  none \\
HCP97   &  hexagonal  &     no  & 5003   & 3.75      & $\mathbf{10^{21}}$  & $400$  & 0.00          & 0 & 90  &  none \\
HCP98   &  hexagonal  &     no  & 5003   & 3.75      & $\mathbf{10^{22}}$  & $860$  & 0.00          & 0 &  $\mathbf{180}$ &  none \\
HCP99   &  hexagonal  &     no  & $\mathbf{3985}$   & 3.75      & $\mathbf{10^{22}}$  & $860$  & 0.00          & 0 & $\mathbf{365}$  &  none \\
HCP100  &  hexagonal  &     no  & $\mathbf{3011}$   & 3.75      & $\mathbf{10^{22}}$  & $860$  & 0.00          & 0 & $\mathbf{730}$  &  none \\
\hline
HCP101   &  hexagonal  &     no  & 5003   & 4.00      & $\mathbf{10^{16}}$  & $8.4$   & 0.00          & 0 & 90  &   none  \\
HCP102   &  hexagonal  &     no  & 5003   & 4.00      & $\mathbf{10^{17}}$  & $18$  & 0.00          & 0 & 90  &   none \\
HCP103   &  hexagonal  &     no  & 5003   & 4.00      & $\mathbf{10^{18}}$  & $39$  & 0.00          & 0 & 90  &   none \\
HCP104   &  hexagonal  &     no  & 5003   & 4.00      & $\mathbf{10^{19}}$  & $84$  & 0.00          & 0 & $\mathbf{180}$  &   none \\
HCP105   &  hexagonal  &     no  & $\mathbf{3985}$   & 4.00      & $\mathbf{10^{19}}$  & $84$  & 0.00          & 0 & $\mathbf{365}$  &  none \\
HCP106   &  hexagonal  &     no  & $\mathbf{3011}$   & 4.00      & $\mathbf{10^{19}}$  & $84$  & 0.00          & 0 & $\mathbf{730}$  &  none \\
HCP107   &  hexagonal  &     no  & 5003   & 4.00      & $\mathbf{10^{20}}$  & $180$  & 0.00          & 0 & 90  &  none \\
HCP108   &  hexagonal  &     no  & 5003   & 4.00      & $\mathbf{10^{21}}$  & $390$  & 0.00          & 0 & 90  &  none \\
HCP109   &  hexagonal  &     no  & 5003   & 4.00      & $\mathbf{10^{22}}$  & $840$  & 0.00          & 0 &  $\mathbf{180}$ &  none \\
HCP110   &  hexagonal  &     no  & $\mathbf{3985}$   & 4.00      & $\mathbf{10^{22}}$  & $840$  & 0.00          & 0 & $\mathbf{365}$  &  none \\
HCP111   &  hexagonal  &     no  & $\mathbf{3011}$   & 4.00      & $\mathbf{10^{22}}$  & $840$  & 0.00          & 0 & $\mathbf{730}$  &  none \\
\hline
HCP112   &  hexagonal  &     no  & 5003   & 2.00      & $10^{22}$  & $1100$   & $\mathbf{0.001}$         & 0 & 90  &  full  \\
HCP113   &  hexagonal  &     no  & 5003   & 2.00      & $10^{22}$  & $1100$   & $\mathbf{0.005}$         & 0 & 90  &  full  \\
HCP114   &  hexagonal  &     no  & 5003   & 2.00      & $10^{22}$  & $1100$   & $\mathbf{0.010}$        & 0 & 90  &  full  \\
HCP115   &  hexagonal  &     no  & 5003   & 2.00      & $10^{22}$  & $1100$   & $\mathbf{0.020}$         & 0 & 90  &  full  \\
HCP116   &  hexagonal  &     no  & 5003   & 2.00      & $10^{22}$  & $1100$   & $\mathbf{0.050}$         & 0 & 90  &  full  \\
HCP117   &  hexagonal  &     no  & 5003   & 2.00      & $10^{22}$  & $1100$   & $\mathbf{0.100}$         & 0 & 90  &  full  \\
HCP118   &  hexagonal  &     no  & 5003   & 2.00      & $10^{22}$  & $1100$   & $\mathbf{0.150}$         & 0 & 90  &  full  \\
HCP119   &  hexagonal  &     no  & 5003   & 2.00      & $10^{22}$  & $1100$   & $\mathbf{0.200}$         & 0 & 90  &  full  \\
\hline
HCP120    &  hexagonal  &     no  & 5003   & 4.00      & $10^{22}$  & $840$   &  $\mathbf{0.001}$        & 0 & 90  &   none  \\
HCP121   &  hexagonal  &     no  & 5003   & 4.00      & $10^{22}$  & $840$   &  $\mathbf{0.005}$         & 0 & 90  &   none  \\
HCP122   &  hexagonal  &     no  & 5003   & 4.00      & $10^{22}$  & $840$   &  $\mathbf{0.010}$         & 0 & 90  &   none  \\
HCP123   &  hexagonal  &     no  & 5003   & 4.00      & $10^{22}$  & $840$   &  $\mathbf{0.020}$          & 0 & 90  &   none  \\
HCP124   &  hexagonal  &     no  & 5003   & 4.00      & $10^{22}$  & $840$   &  $\mathbf{0.050}$         & 0 & 90  &   none  \\
HCP125   &  hexagonal  &     no  & 5003   & 4.00      & $10^{22}$  & $840$   &  $\mathbf{0.080}$         & 0 & 90  &   none  \\
HCP126   &  hexagonal  &     no  & 5003   & 4.00      & $10^{22}$  & $840$   &  $\mathbf{0.090}$          & 0 & 90  &   none  \\
HCP127   &  hexagonal  &     no  & 5003   & 4.00      & $10^{22}$  & $840$   &  $\mathbf{0.100}$         & 0 & 90  &   none  \\
HCP128   &  hexagonal  &     no  & 5003   & 4.00      & $10^{22}$  & $840$   &  $\mathbf{0.110}$         & 0 & 90  &   none \\
HCP129   &  hexagonal  &     no  & 5003   & 4.00      & $10^{22}$  & $840$   &  $\mathbf{0.120}$         & 0 & 90  &   none  \\
HCP130   &  hexagonal  &     no  & 5003   & 4.00      & $10^{22}$  & $840$   &  $\mathbf{0.130}$         & 0 & 90  &  full  \\
HCP131   &  hexagonal  &     no  & 5003   & 4.00      & $10^{22}$  & $840$   &  $\mathbf{0.140}$         & 0 & 90  &  full  \\
HCP132   &  hexagonal  &     no  & 5003   & 4.00      & $10^{22}$  & $840$   &  $\mathbf{0.150}$         & 0 & 90  &  full  \\
HCP133   &  hexagonal  &     no  & 5003   & 4.00      & $10^{22}$  & $840$   &  $\mathbf{0.200}$         & 0 & 90  &  full  \\
HCP134    &  hexagonal &        no  & 5003   &  2.60     & $1.0 \times 10^{22}$  & $1000$   & 0.00         & 0 & 90  &  full  \\
\hline
RandDiff1   &  random  &        yes  & 5000   &  $\mathbf{2.50}$     & $8.9 \times 10^{21}$  & $1000$   & 0.00         & 0 & 90  &  full  \\
RandDiff2   &  random  &        yes  & 5000   &  $\mathbf{2.60}$       & $9.3 \times 10^{21}$  & $1000$   & 0.00         & 0 & 90  &  full  \\
RandDiff3   &  random  &        yes  & 5000   &  $\mathbf{2.75}$      & $9.8 \times 10^{21}$  & $1000$   & 0.00         & 0 & 90  &  full  \\
RandDiff4   &  random  &        yes  & 5000   &  $\mathbf{2.90}$      & $1.0 \times 10^{22}$  & $1000$   & 0.00         & 0 & 90  &  full  \\
RandDiff5   &  random  &        yes  & 5000   &  $\mathbf{3.00}$      & $1.1 \times 10^{22}$  & $1000$   & 0.00         & 0 & 90  &  full  \\
RandDiff6   &  random  &        yes  & 5000   &  $\mathbf{3.10}$      & $1.1 \times 10^{22}$  & $1000$   & 0.00         & 0 & 90  &  full  \\
RandDiff7   &  random  &        yes  & 5000   &  $\mathbf{3.20}$      & $1.1 \times 10^{22}$  & $1000$   & 0.00         & 0 & 90  &   mantle  \\
RandDiff8   &  random  &        yes  & 5000   &  $\mathbf{3.50}$      & $1.2 \times 10^{22}$  & $1000$   & 0.00         & 0 & 90  &   mantle  \\
RandDiff9   &  random  &        yes  & 5000   &  $\mathbf{3.80}$      & $1.4 \times 10^{22}$  & $1000$   & 0.00         & 0 & 90  &   mantle  \\
RandDiff10   &  random  &        yes  & 5000   & $\mathbf{3.90}$      & $1.4 \times 10^{22}$  & $1000$   & 0.00         & 0 & 90  &   mantle  \\
RandDiff11   &  random  &        yes  & 5000   & $\mathbf{4.00}$      & $1.4 \times 10^{22}$  & $1000$   & 0.00         & 0 & 90  &   mantle  \\
\hline
RandDiff12   &  random  &        yes  & 5000   &  $\mathbf{2.50}$     & $8.9 \times 10^{21}$  & $1000$   & 0.00         & 1 & 90  &  full  \\
RandDiff13   &  random  &        yes  & 5000   &  $\mathbf{2.60}$       & $9.3 \times 10^{21}$  & $1000$   & 0.00       & 1 & 90  &  full  \\
RandDiff14   &  random  &        yes  & 5000   &  $\mathbf{2.75}$      & $9.8 \times 10^{21}$  & $1000$   & 0.00        & 1 & 90  &  full  \\
RandDiff15   &  random  &        yes  & 5000   &  $\mathbf{2.90}$      & $1.0 \times 10^{22}$  & $1000$   & 0.00        & 1 & 90  &  full  \\
RandDiff16   &  random  &        yes  & 5000   &  $\mathbf{3.00}$      & $1.1 \times 10^{22}$  & $1000$   & 0.00        & 1 & 90  &  full  \\
RandDiff17   &  random  &        yes  & 5000   &  $\mathbf{3.10}$      & $1.1 \times 10^{22}$  & $1000$   & 0.00        & 1 & 90  &  full  \\
RandDiff18   &  random  &        yes  & 5000   &  $\mathbf{3.20}$      & $1.1 \times 10^{22}$  & $1000$   & 0.00        & 1 & 90  &   mantle  \\
RandDiff19   &  random  &        yes  & 5000   &  $\mathbf{3.50}$      & $1.2 \times 10^{22}$  & $1000$   & 0.00        & 1 & 90  &   mantle  \\
RandDiff20   &  random  &        yes  & 5000   &  $\mathbf{3.80}$      & $1.4 \times 10^{22}$  & $1000$   & 0.00        & 1 & 90  &   mantle  \\
RandDiff21   &  random  &        yes  & 5000   & $\mathbf{3.90}$      & $1.4 \times 10^{22}$  & $1000$   & 0.00         & 1 & 90  &   mantle  \\
RandDiff22   &  random  &        yes  & 5000   & $\mathbf{4.00}$      & $1.4 \times 10^{22}$  & $1000$   & 0.00         & 1 & 90  &   mantle  \\
RandDiff23   &  random  &        yes  & 5000   &  4.20     & $1.50 \times 10^{22}$ & $1000$   & 0.00         & 1 & 90  &  mantle  \\
RandDiff24   &  random  &        yes  & 5000   &  4.40     & $1.56 \times 10^{22}$ & $1000$   & 0.00         & 1 & 90  &  none  \\
RandDiff25   &  random  &        yes  & 5000   &  4.60     & $1.64 \times 10^{22}$ & $1000$   & 0.00         & 1 & 90  &  none  \\
\hline
RandDiff26     &  random  &        yes  & 5000   &  2.75     & $9.8 \times 10^{21}$  & $1000$   & 0.00         & 2 & 90  &  full  \\
RandDiff27     &  random  &        yes  & 5000   &  2.75     & $9.8 \times 10^{21}$  & $1000$   & 0.01         & 0 & 90  &  full    \\
RandDiff28     &  random  &        yes  & 5000   &  2.75     & $9.8 \times 10^{21}$  & $1000$   & 0.01         & 1 & 90  &  full    \\
RandDiff29    &  random  &        yes  & 5000   &  3.00     & $1.1 \times 10^{22}$  & $1000$   & 0.01         & 0 & 90  &  full   \\
RandDiff30    &  random  &        yes  & 5000   &  3.00     & $1.1 \times 10^{22}$  & $1000$   & 0.01         & 1 & 90  &  full   \\
RandDiff31    &  random  &        yes  & 5000   &  3.50     & $1.2 \times 10^{22}$  & $1000$   & 0.01         & 0 & 90  &  full    \\
RandDiff32    &  random  &        yes  & 5000   &  3.50     & $1.2 \times 10^{22}$  & $1000$   & 0.01         & 1 & 90  &  full    \\
RandDiff33    &  random  &        yes  & 5000   &  3.50     & $1.2 \times 10^{22}$  & $1000$   & 0.01         & 2 & 90  &  full    \\
RandDiff34   &  random  &        yes  & 5000   &  4.00     & $1.4 \times 10^{22}$  & $1000$   & 0.01         & 0 & 90  &  mantle  \\
RandDiff35   &  random  &        yes  & 5000   &  4.00     & $1.4 \times 10^{22}$  & $1000$   & 0.01         & 1 & 90  &  mantle  \\
RandDiff36   &  random  &        yes  & 5000   &  4.00     & $1.4 \times 10^{22}$  & $1000$   & 0.02         & 0 & 90  &  full  \\
RandDiff37   &  random  &        yes  & 5000   &  4.00     & $1.4 \times 10^{22}$  & $1000$   & 0.02         & 1 & 90  &  full  \\
\hline
HCPDiff1 & hexagonal & yes & 5003 & $\mathbf{2.00}$ &  $7.41 \times 10^{21}$ & 1000 & 0.00 & 1 & 90 & full   \\
HCPDiff2 & hexagonal & yes & 5003 & $\mathbf{2.10}$ &  $7.78 \times 10^{21}$ & 1000 & 0.00 & 1 & 90 & mantle \\
HCPDiff3 & hexagonal & yes & 5003 & $\mathbf{2.20}$ &  $8.15 \times 10^{21}$ & 1000 & 0.00 & 1 & 90 & mantle \\
HCPDiff4 & hexagonal & yes & 5003 & $\mathbf{2.40}$ &  $7.99 \times 10^{21}$ & 965 & 0.00 & 0 & 90 & mantle  \\ 
HCPDiff5 & hexagonal & yes & 5003 & $\mathbf{2.50}$ &  $8.33 \times 10^{21}$ & 965 & 0.00 & 0 & 90 & mantle  \\ 
HCPDiff6 & hexagonal & yes & 5003 & $\mathbf{2.60}$ &  $8.64 \times 10^{21}$ & 965 & 0.00 & 0 & 90 & mantle  \\
HCPDiff7 & hexagonal & yes & 5003 & $\mathbf{2.75}$ &  $9.15 \times 10^{21}$ & 965 & 0.00 & 0 & 90 & mantle  \\
HCPDiff8 & hexagonal & yes & 5003 & $\mathbf{2.80}$ &  $1.04 \times 10^{22}$ & 1000 & 0.00 & 1 & 90 & mantle \\
HCPDiff9 & hexagonal & yes & 5003 & $\mathbf{3.00}$ &  $1.11 \times 10^{22}$ & 1000 & 0.00 & 1 & 90 & mantle \\
HCPDiff10 & hexagonal & yes & 5003 & $\mathbf{3.20}$ & $1.18 \times 10^{22}$ & 1000 & 0.00 & 1 & 90 & none  \\
HCPDiff11 & hexagonal & yes & 5003 & $\mathbf{3.40}$ & $1.26 \times 10^{22}$ & 1000 & 0.00 & 1 & 90 & none  \\
\end{longtable}
\end{center}

\label{lastpage}
\end{document}